\documentclass[useAMS,usenatbib,usegraphicx]{mn2e}
\usepackage{times}
\input{psfig.sty}

\def\gs{\mathrel{\raise0.35ex\hbox{$\scriptstyle >$}\kern-0.6em
\lower0.40ex\hbox{{$\scriptstyle \sim$}}}}
\def\ls{\mathrel{\raise0.35ex\hbox{$\scriptstyle <$}\kern-0.6em
\lower0.40ex\hbox{{$\scriptstyle \sim$}}}}

\title[A robust sample of submillimetre galaxies]
      {A robust sample of submillimetre galaxies: constraints on the prevalence
       of dusty, high-redshift starbursts}
\author[Ivison et al.]
       {R.\ J.\ Ivison,$^{\! 1,2}$
        Ian Smail,$\! ^3$ J.\ S.\ Dunlop,$\! ^2$ 
        T.\ R.\ Greve,$^{\! 4}$ A.\ M.\ Swinbank,$\! ^3$ \and
        J.\ A.\ Stevens,$^{\! 1,5}$ A.\ M.\ J.\ Mortier,$\! ^6$
        S.\ Serjeant,$\! ^6$ T.\ A.\ Targett,$\! ^2$ F.\ Bertoldi,$\! ^7$
        \and A.\ W.\ Blain$^4$ and S.\ C.\ Chapman$^4$
        \vspace*{1mm}\\
        $^1$ UK Astronomy Technology Centre, Royal Observatory, Blackford Hill,
             Edinburgh EH9 3HJ\\
        $^2$ Institute for Astronomy, University of Edinburgh, Blackford Hill,
             Edinburgh EH9 3HJ\\
        $^3$ Institute for Computational Cosmology, University of Durham,
             South Road, Durham DH1 3LE\\
        $^4$ Astronomy Department, California Institute of
             Technology, Pasadena, CA 91125, USA\\
        $^5$ Centre for Astrophysics Research, Science and Technology
             Research Centre,
             University of Hertfordshire, College Lane, Hatfield AL10 9AB\\
        $^6$ Centre for Astrophysics \& Planetary Science, School
             of Physical Sciences, University of Kent, Canterbury, Kent
             CT2 7NR\\
        $^7$ Radioastronomisches Institut der Universit{\"a}t Bonn, Auf dem
             H{\"u}gel 71, D-53121 Bonn, Germany
}

\date{Accepted ... ; Received ... ; in original form ...}

\pagerange{000--000}

\begin{document}

\maketitle

\begin{abstract} The modest significance of most sources detected in
current (sub)millimetre surveys can potentially compromise some
analyses due to the inclusion of spurious sources in catalogues
typically selected at $\ge$3.0--3.5$\sigma$. Here, we develop and
apply a dual-survey extraction technique to SCUBA and MAMBO images of
the Lockman Hole. Cut above 5$\sigma$, our catalogue of submillimetre
galaxies (SMGs) is more robust than previous samples, with a reduced
likelihood of real, but faint SMGs (beneath and around the confusion
limit) entering via superposition with noise. Our selection technique
yields 19 SMGs in an effective area of 165\,arcmin$^2$, of which we
expect at most two to be due to chance superposition of SCUBA and
MAMBO noise peaks. The effective flux limit of the survey
($\sim$4\,mJy at $\sim$1\,mm) is well matched to our deep 1.4-GHz
image ($\sigma=\rm 4.6\,\mu$Jy beam$^{-1}$). The former is sensitive
to luminous, dusty galaxies at extreme redshifts whilst the latter
probes the $z\ls\rm 3$ regime. A high fraction of our robust SMGs
($\sim$80 per cent) have radio counterparts which, given the
$\sim$10-per-cent contamination by spurious sources, suggests that
very distant SMGs ($z\gg\rm 3$) are unlikely to make up more than
$\sim$10 per cent of the bright SMG population. This implies that
almost all of the $S_{\rm 1mm}\gs\rm 4$\,mJy SMG population is
amenable to study via the deepest current radio imaging. We use these
radio counterparts to provide an empirical calibration of the
positional uncertainty in SMG catalogues. We then go on to outline the
acquisition of redshifts for radio-identified SMGs, from sample
selection in the submillimetre, to counterpart selection in the radio
and optical/infrared, to slit placement on spectrograph masks. We
determine a median of $z=\rm 2.05\pm 0.41$ from a sample of six secure
redshifts for unambigious radio-identified submillimetre sources and
$z=\rm 2.14\pm 0.27$ when we include submillimetre sources with
multiple radio counterparts and/or less reliable redshifts.  These
figures are consistent with previous estimates, suggesting that our
knowledge of the median redshift of bright SMGs population has not
been biased by the low significance of the source catalogues employed.
\end{abstract}

\begin{keywords}
   galaxies: starburst
-- galaxies: formation
-- cosmology: observations
-- cosmology: early Universe
\end{keywords}

\section{Introduction}

%
%
\setcounter{figure}{0}
\begin{figure*}
\centerline{\psfig{file=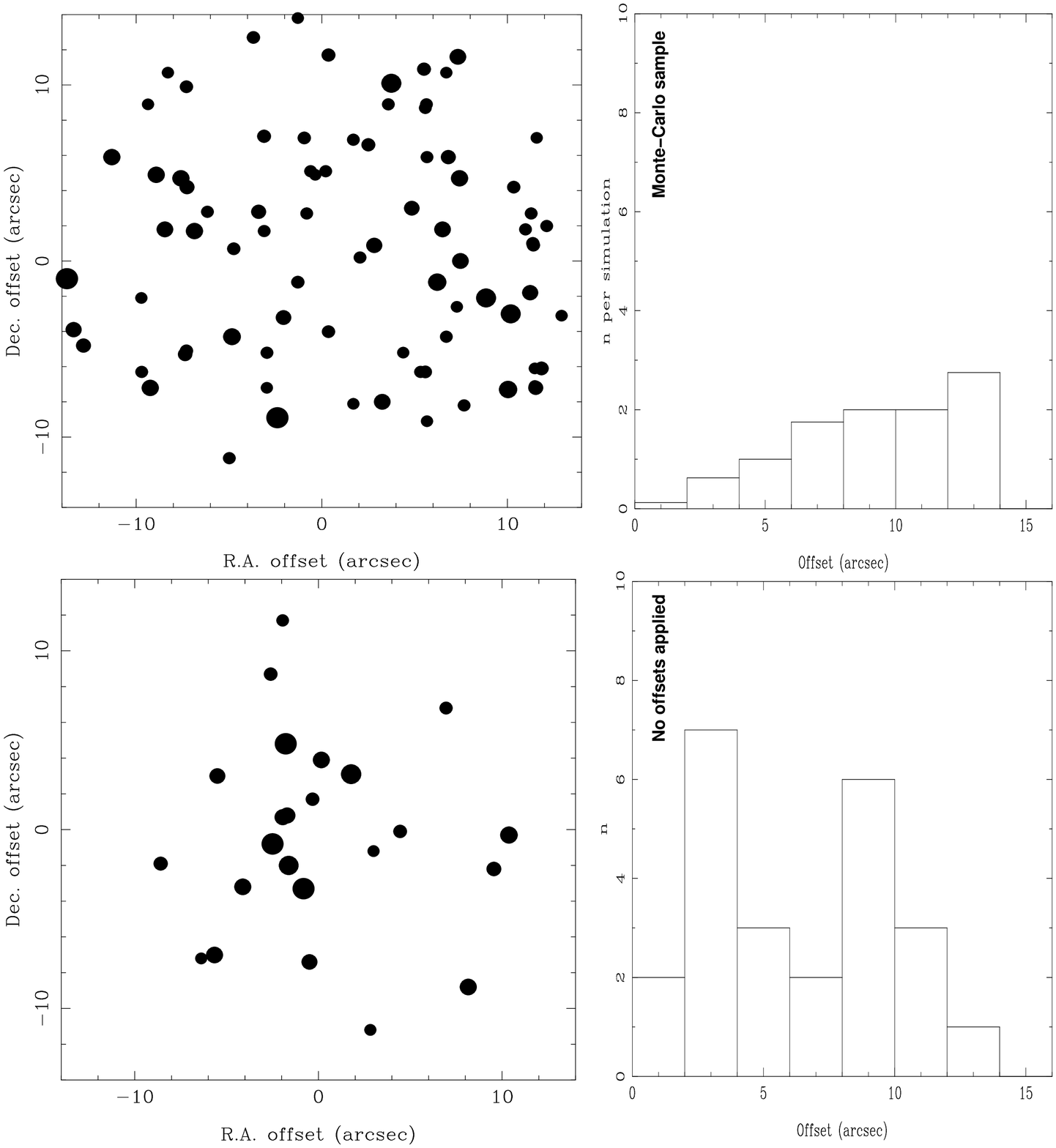,angle=0,width=3.1in}
\hspace*{10mm}
\psfig{file=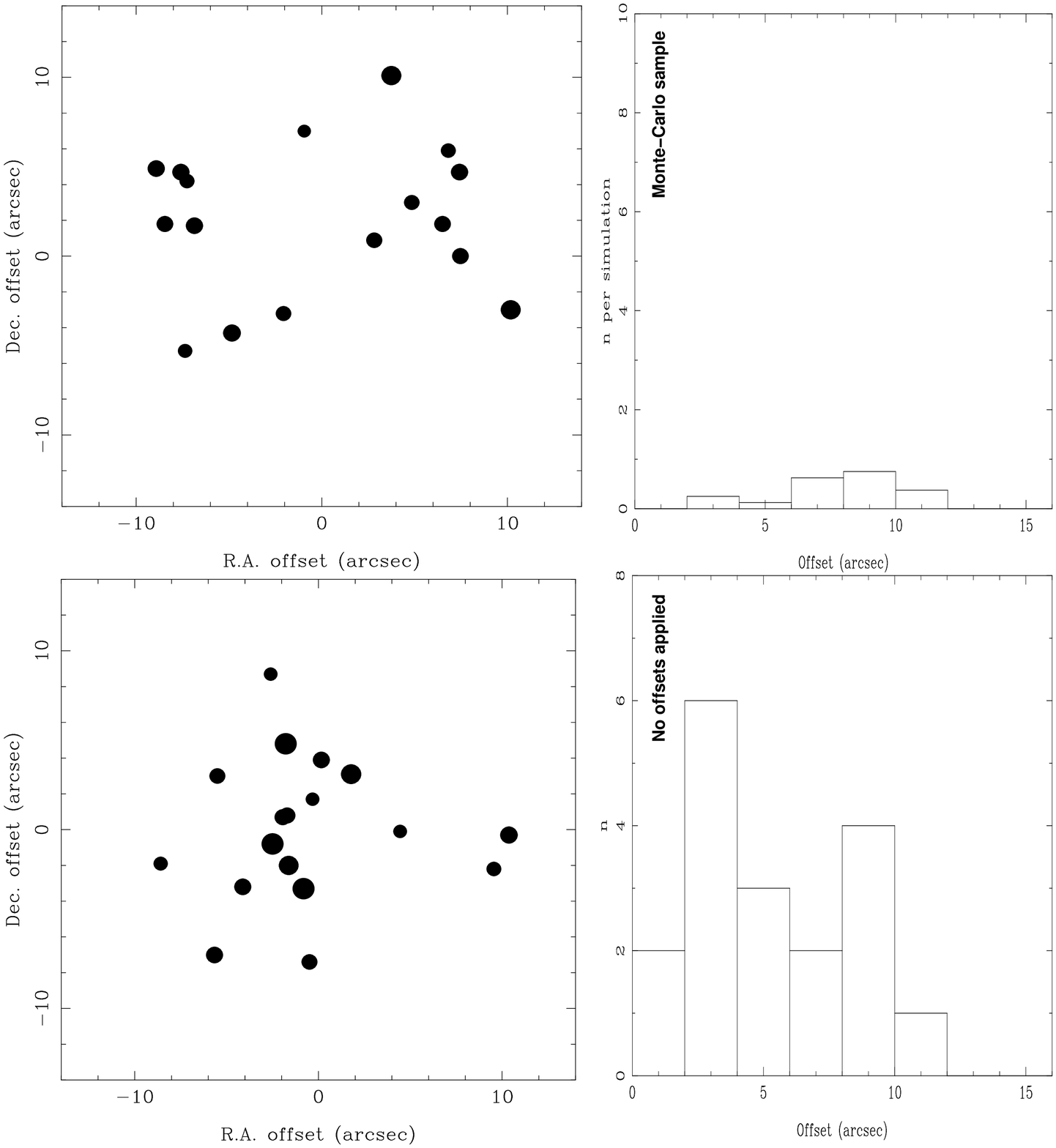,angle=0,width=3.1in}}
\vspace{-0.3cm}
\noindent{\small\addtolength{\baselineskip}{-3pt}}
\caption{{\em Left-hand panels:} Positional offsets between SCUBA and
MAMBO peaks for spurious ({\em top}) and real ({\em bottom}) catalogues
with a combined significance above 4.5$\sigma$.  Larger symbols
represent a higher combined SNR. Spurious sources were generated by
offsetting MAMBO catalogue positions by $\pm$30$''$ and $\pm$45$''$ (in
R.A.\ and Dec.).  Histograms show the radial offset ($r = \sqrt (\Delta
\alpha^2 + \Delta \delta^2)$) between SCUBA and MAMBO peaks for
spurious ({\em top}, normalised to one simulation) and real ({\em
bottom}) catalogues.  {\em Right-hand panels:} The same plots, after
application of the selection filters described in \S2.2.}
\label{offset_before}
\end{figure*}

Surveys with bolometer arrays at millimetre (mm) and submillimetre (submm)
wavelengths are potentially sensitive to dusty objects at extreme redshifts,
galaxies that drop out of surveys at shorter and longer wavelengths due to
obscuration and unfavourable $K$ corrections. The first cosmological surveys
using SCUBA (Holland et al.\ 1999) and MAMBO (Kreysa et al.\ 1998) quickly and
radically changed the accepted picture of galaxy formation and evolution,
moving away from the optocentric view of the last century. The discovery of
so-called `SCUBA galaxies' (Smail, Ivison \& Blain 1997) was greeted with
surprise due to the remarkable evolution in the dusty, starburst galaxy
population implied by such a large source density at the flux levels accessible
to the first generation of bolometer arrays (Blain et al.\ 1999). Excitement
was replaced by pessimism with the first efforts to study SMGs at optical and
infrared (IR) wavelengths: early reports, backed up with a study in the Hubble
Deep Field North by Hughes et al.\ (1998), suggested that the majority of the
submm population had no plausible optical counterparts. Attention was diverted
to various redshift engines and broadband photometric techniques (e.g.\
Townsend et al.\ 2001; Aretxaga et al.\ 2003; Wiklind 2003). As a result, only
a handful of detailed studies were attempted, often for extreme and possibly
unrepresentative galaxies (e.g.\ Knudsen et al.\ 2004).

Recent progress has largely been the result of radio imaging of submm
survey fields. Early radio follow-up detected roughly half of the
submm sources observed (Smail et al.\ 2000; Ivison et al.\ 2002 ---
hereafter I02), with an astrometric precision of $\sim$0.3$''$ and,
combined with the submm flux density, provide a rough estimate of
redshift (Carilli \& Yun 1999). Radio data also enabled some
refinement of submm samples (I02), increasing the detection fraction
to two thirds of SMGs at 0.85-mm flux density levels in excess of
$\sim$5\,mJy. With positions in hand, these bright SMGs were found to
be a diverse population --- some quasar-like, with broad lines and
X-ray detections (e.g.\ Ivison et al.\ 1998), some morphologically
complex (e.g.\ Ivison et al.\ 2000; cf.\ Downes \& Solomon 2003;
Smail, Smith \& Ivison 2005), some extremely red (e.g.\ Smail et al.\
1999; Gear et al.\ 2000; I02; Webb et al.\ 2003b; Dunlop et al.\
2004), some with the unmistakable signatures of obscured active nuclei
and/or superwinds (e.g.\ Smail et al.\ 2003).

Spectroscopic redshifts have been difficult to determine. The first
survey based on a submm/radio sample was undertaken by Chapman et al.\
(2003, 2005 --- hereafter C03, C05): the median redshift was found to
be $\sim$2.2 for $S_{\rm 0.85mm} \ge \rm 5$-mJy galaxies selected
using SCUBA and pinpointed at 1.4\,GHz. The accurate redshifts
reported by C03 and C05 facilitated the first systematic measurements
of molecular gas mass for SMGs ($\sim$10$^{11}$\,M$_{\odot}$) via
observations of CO (Neri et al.\ 2003; Greve et al.\ 2005), as well as
constraints on gas reservoir size and dynamical mass (Tacconi et al.\
2005). The data suggest SMGs are massive systems and provide some of
the strongest tests of galaxy-formation models to date (Greve et al.\
2005).

In spite of this progress, a detailed understanding of SMGs remains a
distant goal. Confusion currently limits our investigations to the
brightest SMGs (although surveys through lensing clusters have
provided a handful of sources more typical of the faint population
that dominates the cosmic background --- Smail et al.\ 2002; Kneib et
al.\ 2004; Borys et al.\ 2004). We must also recall that selection
biases have potentially skewed our understanding: around half of all
known SMGs remain undetected in the radio (due simply to the lack of
sufficiently deep radio data, which do not benefit from the same $K$
correction as submm data) and the radio-undetected fraction remains
largely untargeted by existing spectroscopic campaigns. These is also
only limited coverage of red and IR wavelengths in spectroscopic
surveys.

Here, we present a robust sample of bright SMGs selected using SCUBA
and MAMBO in one of the `8-mJy Survey' regions: the Lockman Hole (see
Scott et al.\ 2002; Fox et al.\ 2002; I02; Greve et al.\ 2004; Mortier
et al.\ 2005). Our goal is to provide a bright sample which we would
expect to detect in well-matched radio imaging ($\sigma_{\rm
0.85mm}/\sigma_{\rm 1.4GHz} \sim\rm 500$) whilst minimising, so far as
is practicable, the possibility that sources are spurious or
anamalously bright. We may thus determine the true fraction of radio
drop-outs amongst SMGs (potentially lying at very high redshift,
$z\gg\rm 3$), as well as practical information such as the intrinsic
positional uncertainty for SMGs in the absence of radio/IR
counterparts.

Throughout we adopt a cosmology, with $\Omega_m=0.3$, $\Omega_\Lambda=0.7$ and
$H_0=70$\,km\,${\rm s^{-1}}$\,Mpc$^{-1}$.

\section{Sample selection}

\subsection{Strategy}

Existing surveys have typically employed a SNR threshold of
3.0--3.5. At these SNRs, false detections are dominated by `flux
boosting' (\S2.2), possibly at the 10--40 per cent level (Scott et
al.\ 2002; Laurent et al.\ 2005). Our goal is to provide a highly
reliable submm source catalogue, free from concerns about
contamination by spurious or artificially bright sources. This issue
has limited our ability to address the true recovery fraction in the
radio, and hence the corrections that must be made to the redshift
distributions that are used to determine star-formation histories and
galaxy-formation models.

%
%
\setcounter{figure}{1}
\begin{figure}
\centerline{\psfig{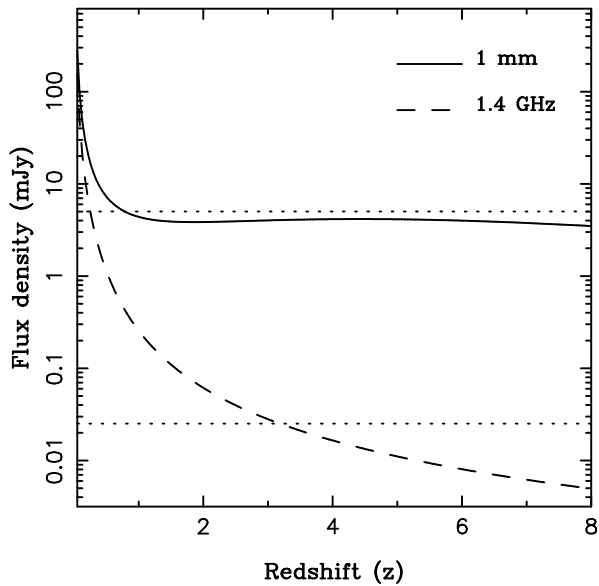}}
\vspace{-0.3cm}
\noindent{\small\addtolength{\baselineskip}{-3pt}}
\caption{Flux density versus redshift for an Arp\,220 spectral energy
distribution. An IR luminosity of $\rm 6 \times 10^{12}$\,L$_{\odot}$
is used, where $L_{\rm IR}$ for Arp\,220 is $\rm 1.5 \times
10^{12}$\,L$_{\odot}$. The variation with redshift is shown for flux
densities at 1\,mm and 1.4\,GHz (solid and dotted lines,
respectively), with the 5-$\sigma$ detection limits indicated as
horizontal lines. As is widely known, mm and submm surveys are
sensitive to dusty starbursts out to extreme redshifts ($z\gs\rm 6$)
but the deepest current radio surveys lose sensitivity to these
galaxies at $z\gg\rm 3$. Note that the radio spectral index of
Arp\,220 ($\alpha=\rm -0.3$, where $F_{\nu}\propto \nu^{\alpha}$)
produces a more gentle decline in radio flux density with redshift
than would be expected for starbursts with a more typical spectral
slope ($\alpha\sim\rm -0.7$).}
\label{flux}
\end{figure}

To achieve this we have combined independent submm and mm maps of the
Lockman Hole, constructing a single, reliable catalogue that is
several times larger than would have been realised by simply adopting
a high SNR threshold in the individual submm and mm maps.  Greve et
al.\ (2004) argued that several maps with low signal-to-noise ratio
(SNR) of the same region, with only marginal differences in frequency,
produce several visualisations of essentially the same sky, tracing
the same population of luminous, dust-enshrouded galaxies and we have
adopted the same philosophy in the present work.

\subsection{Practicalities}

Our maps came from the survey of Greve et al.\ (2004) who presented a
1.2-mm map of the Lockman Hole region (as well as data on the ELAIS N2
region), centred on the coordinates mapped at 0.85\,mm by SCUBA in the
`8-mJy Survey' (Scott et al.\ 2002, 2004).  Mortier et al.\ (2005)
have recently present a refined analysis of SCUBA data which lies
within the Lockman Hole MAMBO map; we began with a 2-$\sigma$ MAMBO
catalogue, extracted as described by Greve et al.\ (2004), and looked
for 0.85-mm sources within 14\,arcsec (roughly the area of five beams)
in the Mortier et al.\ (2005) sample, applying a correction for
separation and checking for a combined significance above our initial
threshold, 4.5$\sigma$. We included sources that exceed 4.5$\sigma$ in
either dataset, as long as there is a valid reason why the source is
not seen in the other image. SMGs from Scott et al.\ (2002) were
substituted where blends were evident in the Mortier et al.\
catalogue.

%
%
\setcounter{table}{0}
\begin{table*}
\scriptsize
\caption{Combined sample of MAMBO/SCUBA sources in the Lockman Hole.}
\vspace{0.2cm}
\begin{center}
\begin{tabular}{lcccccccccc}
Name$^a$&\multicolumn{2}{c}{Position at 1.2mm}&$S_{\rm 1.2mm}$&S/N&\multicolumn{2}{c}{Position at 0.85mm}&$S_{\rm 0.85mm}$&S/N&Sep&Final\\
        &$\alpha_{\rm J2000}$&$\delta_{\rm J2000}$&/mJy&&$\alpha_{\rm J2000}$&$\delta_{\rm J2000}$&/mJy&&&S/N\\
        &h m s&$^{\circ}\ '\ ''$&&&h m s&$^{\circ}\ '\ ''$&&&$''$&\\
&&&&&&&&&\\
LH\,1200.001 = LE850.02 = Lock850.03     &10:52:38.3&+57:24:37&4.8\,$\pm$\,0.6&8.0&10:52:38.6&+57:24:38&10.9\,$\pm$\,2.1&5.1&2.1&$>$8\\
LH\,1200.002 = Lock850.32$^b$            &10:52:38.8&+57:23:21&4.1\,$\pm$\,0.6&6.8&10:52:38.6&+57:23:19&5.2\,$\pm$\,2.0 &2.6&3.6&7.6 \\
LH\,1200.003 = LE850.14 = Lock850.04$^c$ &10:52:04.1&+57:26:57&3.6\,$\pm$\,0.6&6.0&10:52:04.2&+57:27:01&10.5\,$\pm$\,2.0&5.2&3.4&$>$8\\
                                         &10:52:03.9&+57:27:10&1.3\,$\pm$\,0.6&2.2&          &         &                &   &   &    \\
LH\,1200.004 = Lock850.02$^d$            &10:52:57.0&+57:21:07&5.7\,$\pm$\,1.0&5.7&          &         &                &   &   &5.7 \\
LH\,1200.005 = LE850.01 = Lock850.01     &10:52:01.3&+57:24:48&3.4\,$\pm$\,0.6&5.7&10:52:01.5&+57:24:43&8.6\,$\pm$\,1.1 &8.0&5.1&$>$8\\
LH\,1200.006 = LE850.16 = Lock850.12     &10:52:27.4&+57:25:15&2.8\,$\pm$\,0.5&5.6&10:52:27.6&+57:25:17&7.0\,$\pm$\,1.6 &4.3&2.6&7.4 \\
LH\,1200.007 = Lock850.27                &10:52:04.7&+57:18:12&3.2\,$\pm$\,0.7&4.6&10:52:03.4&+57:18:12&5.8\,$\pm$\,1.3 &4.3&10.4&6.4\\
LH\,1200.008                             &10:51:41.9&+57:19:51&4.1\,$\pm$\,0.9&4.6&10:51:40.7&+57:19:53&4.6\,$\pm$\,1.6 &2.8&9.8&5.5\\
LH\,1200.009 = Lock850.18                &10:52:27.5&+57:22:20&3.1\,$\pm$\,0.7&4.4&10:52:28.2&+57:22:17&6.3\,$\pm$\,1.8 &3.5&6.3&6.0\\
LH\,1200.010 = LE850.06 = Lock850.14     &10:52:29.9&+57:22:05&2.9\,$\pm$\,0.7&4.1&10:52:30.4&+57:22:13&10.8\,$\pm$\,2.4&4.6&9.5&6.4\\
LH\,1200.011 = LE850.03 = Lock850.17$^e$ &10:51:58.2&+57:17:53&2.9\,$\pm$\,0.7&4.1&10:51:58.3&+57:18:00&5.0\,$\pm$\,1.3 &3.9&7.6&6.0\\
                                         &10:51:57.6&+57:18:05&1.6\,$\pm$\,0.7&2.3&          &         &                &   &   &   \\
LH\,1200.012 = LE850.18 = Lock850.33     &10:51:55.4&+57:23:10&3.3\,$\pm$\,0.8&4.1&10:51:55.9&+57:23:13&4.3\,$\pm$\,1.0 &4.2&5.2&6.2\\
LH\,1200.014 = LE850.08 = Lock850.41     &10:52:00.0&+57:24:24&2.4\,$\pm$\,0.6&4.0&10:51:59.6&+57:24:21&4.8\,$\pm$\,1.1 &4.3&4.9&6.2\\
LH\,1200.017 = Lock850.61                &10:51:21.4&+57:18:40&4.8\,$\pm$\,1.3&3.7&10:51:22.5&+57:18:42&14.6\,$\pm$\,4.1&3.5&8.8&5.3\\
LH\,1200.019                             &10:51:28.3&+57:19:46&4.0\,$\pm$\,1.1&3.6&10:51:27.8&+57:19:47&6.7\,$\pm$\,2.3 &2.9&4.5&5.0\\
LH\,1200.022                             &10:52:03.0&+57:15:46&2.8\,$\pm$\,0.8&3.5&10:52:03.3&+57:15:37&7.3\,$\pm$\,2.2 &3.3&9.1&5.0\\
LH\,1200.042 = LE850.29 = Lock850.09     &10:52:16.0&+57:25:06&1.6\,$\pm$\,0.5&3.2&10:52:16.2&+57:25:05&7.2\,$\pm$\,1.5 &4.7&1.9&6.0\\
LH\,1200.096 = LE850.07 = Lock850.16     &10:51:51.4&+57:26:40&1.6\,$\pm$\,0.6&2.7&10:51:51.4&+57:26:38&6.7\,$\pm$\,1.7 &4.0&4.8&5.1\\
LH\,1200.104 = LE850.27 = Lock850.08     &10:51:53.7&+57:18:39&2.1\,$\pm$\,0.8&2.6&10:51:53.9&+57:18:38&6.3\,$\pm$\,1.2 &5.2&2.1&6.2\\
\end{tabular}
\end{center}

\label{catalogue}
\noindent
Notes:
$a)$ Sources names: LH\,1200.$nnn$ from Greve et al.\ (2004), LE850.$nn$
     from Scott et al.\ (2002) and Lock850.$nn$ from Coppin et al.\
     (in preparation);
$b)$ Noisy region of 0.85-mm 8-mJy Survey map;
$c)$ LH\,1200.213 lies nearby, with coordinates and flux density as shown;
$d)$ 1.2-mm source lies off the 0.85-mm map;
$e)$ LH\,1200.182 lies nearby, with coordinates and flux density as shown;
     also a possible blend at 0.85\,mm.

\end{table*}

%
%
\setcounter{figure}{2}
\begin{figure*}
\centerline{\psfig{file=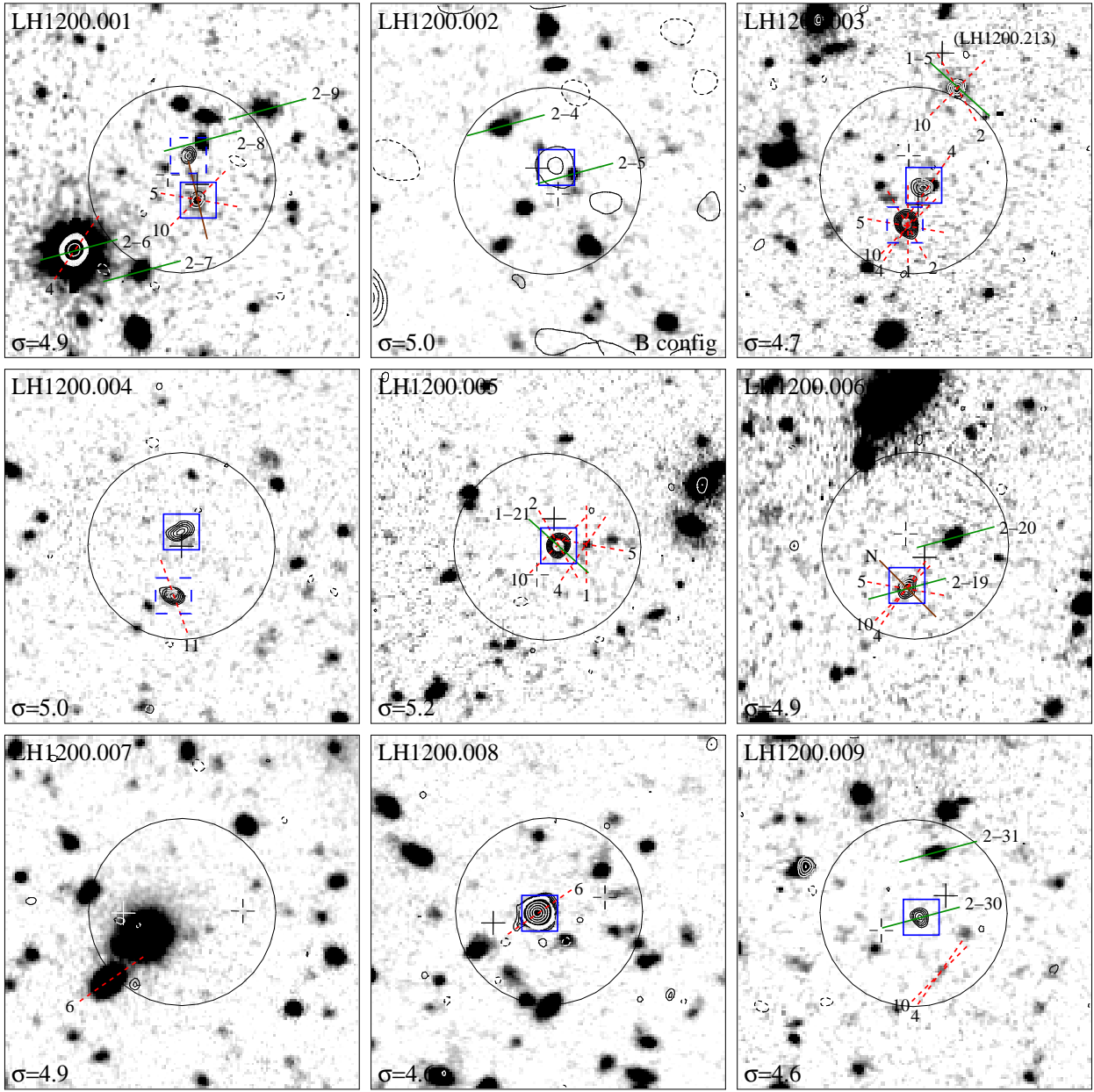,angle=0,width=6.5in}}
\vspace{-0.3cm}
\noindent{\small\addtolength{\baselineskip}{-3pt}}
\caption{Postage stamps (30\,$\times$\,30\,arcsec) of the fields
surrounding the $>$5$\sigma$ SMGs in the Lockman Hole. Optical ($R$)
data are shown as a grayscale upon which 1.4-GHz contours are plotted
at $-$3,3,4,5,6,7,8,9,10,20,30,40,50 $\times$ $\sigma_{1.4}$, where
$\sigma_{1.4}$ has been measured locally and is indicated in the
bottom-left corner of each stamp in units of $\mu$Jy.  Open and solid
crosses mark the nominal centroids of SCUBA and MAMBO galaxies,
respectively; 8-arcsec-radius circles ($\sim$99.9-per-cent positional
confidence --- see \S3.3) mark the average positions; dashed red lines
represent the position of slits on C05's LRIS masks, marked with the
mask number; solid green lines represent slits on the GMOS masks,
marked with the mask and object identification number (format: M--NN);
solid brown lines mark NIRSPEC slit positions; blue squares mark radio
counterparts with $P<\rm 0.05$, solid squares being the most probable
in each case (see \S3.2). N is up; E is to the left.  We have excluded
LH1200.104 as a bright diffraction spike from a nearby star
obliterates any useful information in that region.  }
\label{counterparts}
\end{figure*}

%
%
\setcounter{figure}{2}
\begin{figure*}
\centerline{\psfig{file=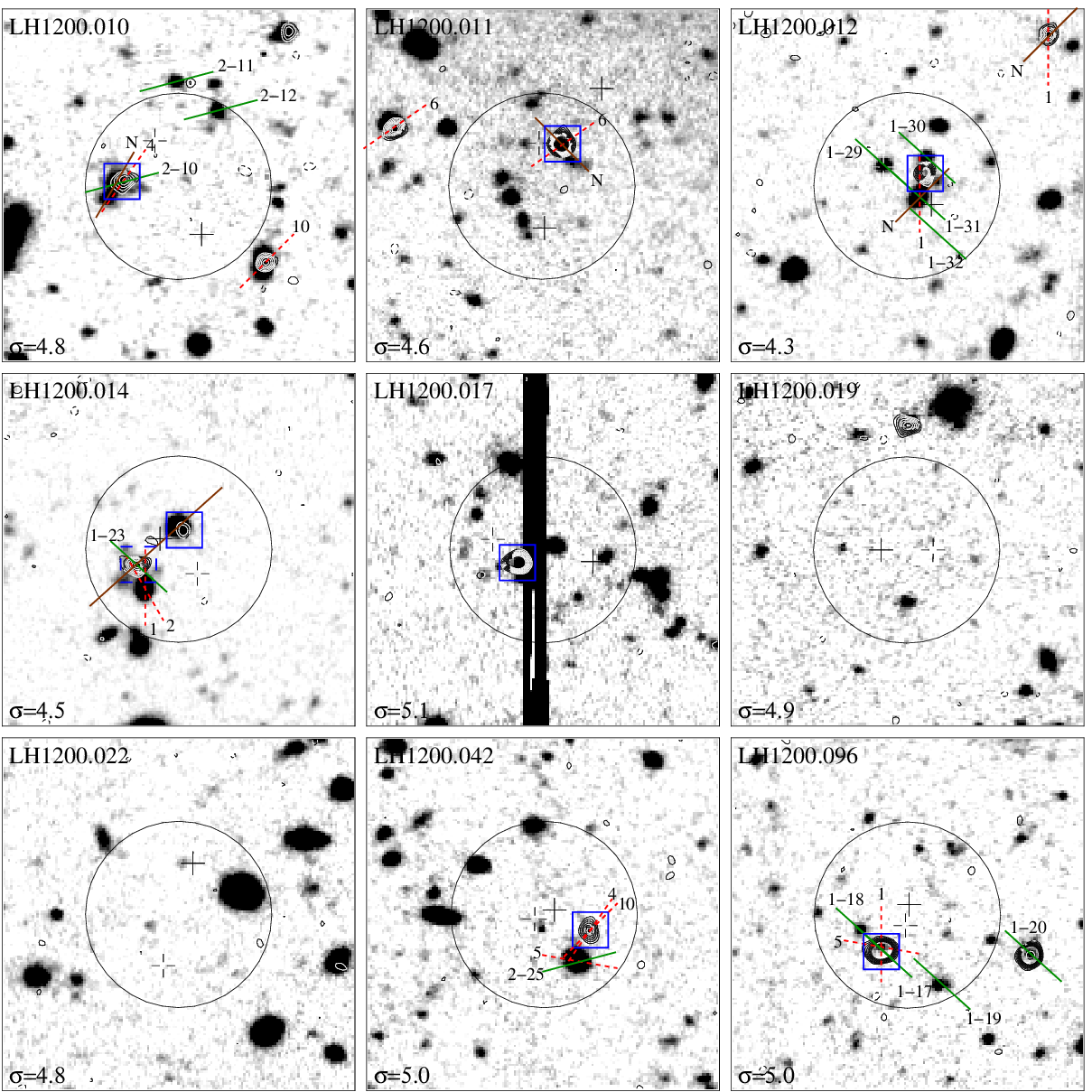,angle=0,width=6.5in}}
\vspace{-0.3cm}
\noindent{\small\addtolength{\baselineskip}{-3pt}}
\caption{continued...}
\label{counterparts2}
\end{figure*}

The noise properties of submm images are not Gaussian due to the
miriad of real, faint sources near the confusion limit. The most
important effect of increasing the SNR threshold should be to
drastically reduce the false detections due to faint SMGs that have
been boosted above the detection threshold by noise --- flux boosting,
see Scott et al.\ (2002) and Greve et al.\ (2004). However, with so
many $\ge$2-$\sigma$ sources in our catalogues, it was important to
investigate how many SMGs may have contaminated our combined sample
due to random coincidence of faint MAMBO and SCUBA peaks. To this end,
we performed simulations, offsetting the MAMBO sample by $\pm$30 and
$\pm$45\,arcsec (in R.A.\ and Dec.). Each simulation typically yielded
ten SMGs --- far too many to hope that a statistically robust
catalogue would emerge from the process. The results of these
simulations are illustrated in Fig.~1. The simulated catalogues
clearly yield a very different distribution of offsets from the
position-matched data, closer to the form expected for randomly
scattered peaks ($n \propto r^2$), and without the expected
concentration of high-SNR pairs at low $r$.

Using the information in these plots we took a number of approaches to
minimise the number of spurious SMGs in our sample. We found that
raising the minimum catalogue threshold to $\ge$2.5$\sigma$ reduced
the number of false detections whilst having no effect on the real
catalogue. Lowering the search radius to 11\,arcsec reduced the number
of false positives in line with the ratio of the respective search
areas, removing only one source from the real catalogue. Insisting
that the higher of the two peaks was $\ge$3.5$\sigma$ removed a
further quarter of the simulated SMGs, whilst reducing the real
catalogue by half that amount. Finally, we increased the combined
theshold to $\ge$5$\sigma$. The effects of this approach on the
SCUBA--MAMBO positional offsets are shown in the right-hand set of
panels of Fig.\ 1. We were left with a catalogue of 19 SMGs, of which
we expect {\it at most} two to be the result of chance superposition
of SCUBA and MAMBO noise peaks. This compares well with the sample of
seven that would have resulted from adopting a $\ge$5$\sigma$
threshold in any one image. We list the resulting 19 sources in
Table~1.

We note that all four of the SCUBA sources detected individually above
5$\sigma$ are also found in the MAMBO image with a SNR of at least
2.0, although one is below 2.5$\sigma$ and has thus been excised from
our final sample. Of the six $\ge$5-$\sigma$ MAMBO sources, all those
that fall within the original 8-mJy Survey region were detected by
SCUBA at $\ge$2.5$\sigma$. LH\,1200.004, which lies off the 8-mJy
Survey SCUBA map, to the East, was subsequently detected by the SHADES
survey (Dunlop 2005; {\tt http://www.roe.ac.uk/ifa/shades}).

%
%
\setcounter{table}{1}
\begin{table*}
\scriptsize
\caption{Radio properties of MAMBO/SCUBA sources in the Lockman Hole.}
\vspace{0.2cm}
\begin{center}
\begin{tabular}{lcccccccl}
Name&\multicolumn{2}{c}{Average (sub)mm position}&\multicolumn{2}{c}{Radio position}       &$S_{\rm 1.4GHz}^a$&Radio-&$P$&Notes\\
    &$\alpha_{\rm J2000}$&$\delta_{\rm J2000}$   &$\alpha_{\rm J2000}$&$\delta_{\rm J2000}$&                &submm &&\\
    &h m s&$^{\circ}\ '\ ''$                     &h m s&$^{\circ}\ '\ ''$                  &/$\mu$Jy        &offset $''$&&\\
&&&&&&&&\\
LH\,1200.001  &10:52:38.46&+57:24:37.4&10:52:38.30&+57:24:35.8&29\,$\pm$\,11 &2.1&0.0237&to SSW\\
              &           &           &10:52:38.39&+57:24:39.5&24\,$\pm$\,9  &2.2&0.0309&to NNW\\
LH\,1200.002  &10:52:38.69&+57:23:19.9&10:52:38.46&+57:23:19.6&45\,$\pm$\,20 &1.9&0.0126&central\\
LH\,1200.003  &10:52:04.15&+57:26:59.1&10:52:04.22&+57:26:55.4&72\,$\pm$\,12 &3.7&0.0213&to S\\
              &           &           &10:52:04.06&+57:26:58.5&36\,$\pm$\,12 &1.9&0.0161&central\\
LH\,1200.004  &10:52:57.00&+57:21:07.0&10:52:57.09&+57:21:02.8&44\,$\pm$\,11 &4.3&0.0437&to S\\
              &           &           &10:52:57.01&+57:21:08.3&48\,$\pm$\,12 &3.7&0.0325&central; extended PA 104$^{\circ}$\\
LH\,1200.005  &10:52:01.37&+57:24:45.6&10:52:01.25&+57:24:45.7&73\,$\pm$\,10 &1.0&0.0025&central\\
LH\,1200.006  &10:52:27.50&+57:25:16.0&10:52:27.58&+57:25:12.4&47\,$\pm$\,10 &3.7&0.0332&to SSE\\
LH\,1200.007  &10:52:04.08&+57:18:12.1&10:52:04.58&+57:18:05.9&18\,$\pm$\,8  &7.4&{\em 0.1303}&to SE\\
LH\,1200.008  &10:51:41.31&+57:19:52.0&10:51:41.43&+57:19:51.9&315\,$\pm$\,12&1.0&0.0004&central; just resolved\\
LH\,1200.009  &10:52:27.84&+57:22:18.5&10:52:27.77&+57:22:18.2&29\,$\pm$\,9  &0.6&0.0031&central\\
LH\,1200.010  &10:52:30.15&+57:22:09.1&10:52:30.73&+57:22:09.5&54\,$\pm$\,14 &4.7&0.0406&to E, resolved\\
LH\,1200.011  &10:51:58.23&+57:17:56.6&10:51:58.02&+57:18:00.2&98\,$\pm$\,12 &4.0&0.0172&to NNW\\
LH\,1200.012  &10:51:55.66&+57:23:11.6&10:51:55.47&+57:23:12.7&47\,$\pm$\,10 &1.9&0.0120&38\,$\pm$\,19$\mu$Jy at 4.9\,GHz\\
LH\,1200.014  &10:51:59.80&+57:24:23.1&10:52:00.26&+57:24:21.7&58\,$\pm$\,12 &4.0&0.0300&to ESE\\
              &           &           &10:51:59.76&+57:24:24.8&23\,$\pm$\,10 &1.7&0.0220&to N\\
LH\,1200.017  &10:51:21.93&+57:18:41.0&10:51:22.20&+57:18:39.9&92\,$\pm$\,12 &2.4&0.0082&to ESE\\
LH\,1200.019  &10:51:28.03&+57:19:47.0&10:51:28.03&+57:19:57.7&58\,$\pm$\,12 &10.7&{\em 0.1020}&to N, outside search area\\
LH\,1200.022  &10:52:03.15&+57:15:41.6&           &           &5$\sigma$ $<$ 25&  &{\em N/A}&\\
LH\,1200.042  &10:52:16.11&+57:25:05.6&10:52:15.63&+57:25:04.2&53\,$\pm$\,12 &4.1&0.0341&to WSW\\
LH\,1200.096  &10:51:51.42&+57:26:39.1&10:51:51.69&+57:26:36.0&135\,$\pm$\,13&3.8&0.0112&to SE, just resolved\\
LH\,1200.104  &10:51:53.82&+57:18:38.6&           &           &5$\sigma$ $<$ 25& &{\em N/A}&\\
\end{tabular}
\end{center}

\noindent
Notes:
$a)$ The noise level is around 4.6\,$\mu$Jy\,beam$^{-1}$; however, flux density
uncertainties are larger unless one assumes the source to be unresolved.

\label{radiocat}
\end{table*}

%
%
\setcounter{figure}{2}
\begin{figure}
\centerline{\psfig{file=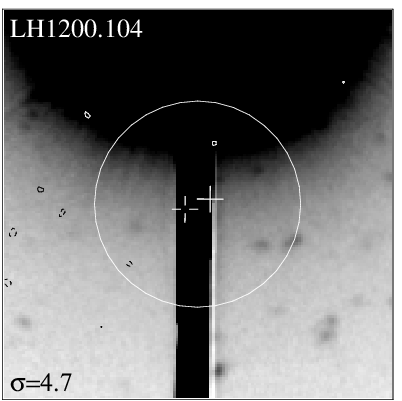,angle=0,width=2.15in}}
\vspace{-0.3cm}
\noindent{\small\addtolength{\baselineskip}{-3pt}}
\caption{continued...}
\label{counterparts3}
\end{figure}

\section{Identification of the SMGs in the radio}

The process of identifying counterparts to SMGs has been refined in a
series of studies (Ivison et al.\ 1998, 2000, 2002, 2004; Smail et
al.\ 1999, 2000; Webb et al.\ 2003a, 2003b; Clements et al.\ 2004;
Borys et al.\ 2004; Pope et al.\ 2005). The most effective methods
employ radio, optical, near-IR and mid-IR imaging, either individually
or in combination. These techniques rely upon identifying the red
rest-frame optical light expected of a dust-enshrouded galaxy and the
synchrotron emission expected of a starburst or radio-loud active
galaxy. Radio sources, EROs and 24-$\mu$m sources are sufficiently
rare that finding either within an SMG error circle can be viewed as a
robust identification in most cases. However, finding an ERO or mid-IR
counterpart does not imply that it is responsible for the submm
emission, merely that it is likely to be associated with the SMG in
some way; for radio counterparts, a more physical link between the
radio and submm emission exists.

Fig.~2 shows how the flux density of a luminous, dusty starburst
varies with redshift at both 1\,mm and 1.4\,GHz, adopting the
starburst spectral energy distribution of Carilli \& Yun (1999). To
identify a $z\le\rm 3$ 5-mJy SMG (the average of the 0.85- and 1.2-mm
flux densities) in a 1.4-GHz image, it is clear that the radio data
must achieve a sensitivity of at least 5\,$\mu$Jy beam$^{-1}$.

Even at this radio sensitivity, the lack of a robust radio
identification could have at least five origins:
\noindent
\begin{enumerate}
\item the SMG could be entirely spurious;
\item the SMG could be real, but flux boosted significantly;
\item the radio/far-IR emission could be significantly larger than the
1.4-GHz synthesised beam (I02; Chapman et al.\ 2004a);
\item the characteristic dust temperature could be low (Chapman et
al.\ 2004b); or
\item the SMG could lie at $z\gg\rm 3$ (Eales et al.\ 2003).
\end{enumerate}

It is difficult, given the quality of existing samples, and the
mis-matched radio/submm datasets, to determine which of these are
important, although I02 showed that (i) probably affects {\em at
least} 15 per cent of a sample selected above 3.5$\sigma$ (cf.\ Pope
et al.\ 2005).

\subsection{Radio data}

Deep, high-resolution, wide-field radio images were obtained at the
National Radio Astronomy Observatory's\footnote{NRAO is operated by
Associated Universities Inc., under a cooperative agreement with the
National Science Foundation.} (NRAO) VLA and employed to pinpoint the
SMGs. The data used here are described in detail by I02. Briefly, the
1.4-GHz image of the Lockman Hole comprises 75\,hr of integration and
reaches a noise level of 4.6\,$\mu$Jy\,beam$^{-1}$, with a
near-circular 1.4-arcsec beam ({\sc fwhm}). We also use an image with
a 4.5-arcsec beam, which incorporates a further 24-hr of
B-configuration data taken for a field 10$'$ to the NE. Figs~3 and 4
show postage stamps around the positions given in Table~2 for the SMGs
in our sample.

\subsection{Mm--radio associations}

A radio source peaking at $\ge$\,4\,$\sigma$ in the 1.4- or 4.5-arcsec
images, with an integrated flux density in excess of 3\,$\sigma$, is
considered a {\it robust} detection. Fainter sources, where the
definition is relaxed to only the integrated flux, were also
catalogued. In the Lockman Hole, the surface density of all radio
sources above this threshold is $\rm 1.9\pm 0.1$\,arcmin$^{-2}$.

For each SMG we have searched for a potential radio (1.4-GHz)
counterpart out to a radius of 8\,arcsec from the mid-point of the
0.85- and 1.2-mm emission. This relatively large search area
(201\,arcsec$^2$ around each source) represents a 2.6-$\sigma$
positional confidence region (see later) and should ensure that few
($\ls 1$ per cent) real associations are missed. As demonstrated by
I02, this search radius can be tolerated without compromising the
statistical significance of genuine associations. Even at the extreme
depths reached by the radio imaging reported here, the cumulative
surface density of radio sources yields only 0.1 source per search
area.

The flux densities and positions of all candidate radio counterparts
are listed in Table~2, along with the search positions. To quantify
the formal significance of each of the potential (sub)mm/radio
associations we have used the method of Downes et al.\ (1986) to
correct the raw Poisson probability, $P$, that a radio source of the
observed flux density could lie at the observed distance from the SMG
for the number of ways that such an apparently significant association
could have been uncovered by chance.

Of the four sources which have more than one potential radio
counterpart, we find that the correct identification is never
statistically obvious.  The formal probability of the second candidate
association occurring by chance is low, $P \le\rm 0.05$. The obvious
interpretations of such multiple statistical associations are either
gravitational lensing (implausible in most of the cases here, unless
the lenses are as obscured as the lensed galaxies) or arise from true
physical associations due to clustering of star-forming objects/AGN at
the source redshift (likely, and already proven in several cases ---
e.g.\ Ledlow et al.\ 2002). Another possibility is that sources with
multiple radio counterparts are boosted into bright submm catalogues
by virtue of comprising multiple, faint, physically unrelated SMGs,
i.e.\ by confusion.

In total, this calculation has yielded statistically robust radio
counterparts for 15 of the 19 SMGs. The plausibility of this figure
can be checked by noting that the ratio of areas inside and outside
the circles in Fig.~3 is 4.5:1. In total, ten random `field' radio
sources are detected robustly in the outer areas. We thus expect only
a handful (at most 2--3) of the counterparts to be spurious,
particularly given that $\mu$Jy radio sources are expected to be
over-dense around SMGs as a result of mergers and/or clustering (Blain
et al.\ 2004).

We find that two SMGs are completely blank in the radio, while a
further two lack robust radio counterparts. Based on the $P$ values
given in Table~2, the most likely candidates for spurious associations
are LH\,1200.007 and .019. LH\,1200.019 and .022 also lack robust {\em
Spitzer} identifications (E.\ Egami, private communication), although
the radio source ($P\sim\rm 0.1$) N of LH\,1200.019 is well detected
at 24\,$\mu$m; LH\,1200.007 has no clear-cut {\em Spitzer}
identification, but two faint 24-$\mu$m sources are detected to the NW
and SE. Finally, LH\,1200.104 has a clear identification in the {\em
Spitzer} imaging described by Ivison et al.\ (in preparation).
LH\,1200.022 is thus the least secure SMG in the sample.

We show in Fig.~5 the distribution of flux ratios for the SMGs in our
robust sample. These demonstrate a real dispersion in 0.85-/1.2-mm and
1.2-mm/1.4-GHz flux ratios, indicative of a broad underlying
distribution of observed spectral energy distributions (SEDs). We can
identify regions of the plot where high- and low-redshift sources
would reside. However, the degeneracy between characteristic dust
temperature and redshift means that variations in dust temperature
within the population may mask any redshift variations. On this basis,
it appears that LH\,1200.042 and .096 may be either particularly hot or
particularly low-redshift SMGs, while LH1200.007 has either unusually
cold dust or lies at a high redshift.

The mean radio flux density observed for our sample is $\rm 89\pm
73$\,$\mu$Jy, taking only the detections and summing flux densities
for those with multiple counterparts. The trend in the most common
spectral index, $S_{\rm 850\mu m}/S_{\rm 1.4GHz}$, determined by C05,
11.1 + 35.2$z$, suggests that the mean redshift of our radio-detected
SMG sample, with $S_{\rm 850\mu m}=\rm 7.4\pm 2.7$\,mJy, is $z\sim\rm
2.2$. The high dispersion in Fig.~5 echoes the findings of C05, who
saw an r.m.s.\ scatter in $S_{\rm 850\mu m}/S_{\rm 1.4GHz}$ of
$\sim$40 for their $z=\rm 1$--4 SMG sample. In fact, the large scatter
provides a plausible explanation for the handful of radio
non-detections amongst our SMGs: we would expect at least 10 per cent
to be scattered below the radio flux density threshold of our survey.

Is there anything which distinguishes the non-radio SMGs from the rest
of the population? Three of the four were detected at $<\rm 4\sigma$ in
the MAMBO survey (LH\,1200.019, .022 and .104). This is interesting
since we would expect the 1.2-mm MAMBO data to be more sensitive than
SCUBA to the most distant starbursts (Fig.~5), and hence to yield more radio
non-detections --- a possible hint that LH\,1200.104, at least, is warm
rather than cold and/or distant. Two of the radio non-detections,
LH\,1200.007 and .022, have the largest mm--submm separations in the
sample and may thus consist of blended, faint sources.

\subsection{Implications for submm positional uncertainty}

Fig.~6 shows a histogram of positional offsets in R.A.\ and Dec.\
between the average of the 0.85- and 1.2-mm positions and the most
likely radio counterparts. A Gaussian fit to the distribution yields a
{\sc fwhm} of 5.2\,$\pm$\,1.2\,arcsec, which translates into a
1-$\sigma$ separation of 2.2\,arcsec between our (sub)mm and radio
positions. The sample can thus be employed to re-calibrate the
rule-of-thumb relationship between positional accuracy, beam size and
SNR (typically, SNR\,$\sim$\,6 here). We must acknowledge a mild
circularity to the logic, given that positional offsets have been used
to calculate $P$, although this may be offset by the lack of
correction for radio sources in the field.

Following Ivison et al.\ (in preparation), the conventional positional
uncertainty, $\sigma_r$, occurs where the distribution of radial
offsets peaks; this is the same as the 1-$\sigma$ separation deduced
earlier. Within this radius we expect to find 39.3 per cent of the
population, with 86.5 and 98.9 per cent within 2\,$\sigma_r$ and
3\,$\sigma_r$ (from 1 -- $e^{-r^2/2\sigma_r^2}$). In the absence of
radio counterparts, it would seem from our analysis that around 39 per
cent of SMGs can be located within a radial distance of
$\sim\theta/\rm SNR$, where $\theta$ is the {\sc fwhm} beam size, in
arcsec). Thus the positional uncertainty ($\sigma_r$) for
3--3.5$\sigma$ SCUBA-selected SMGs is 4--5\,arcsec, cf.\ the rule of
thumb quoted by Hughes et al.\ (1998).

%
%
\setcounter{figure}{3}
\begin{figure*}
\centerline{\psfig{file=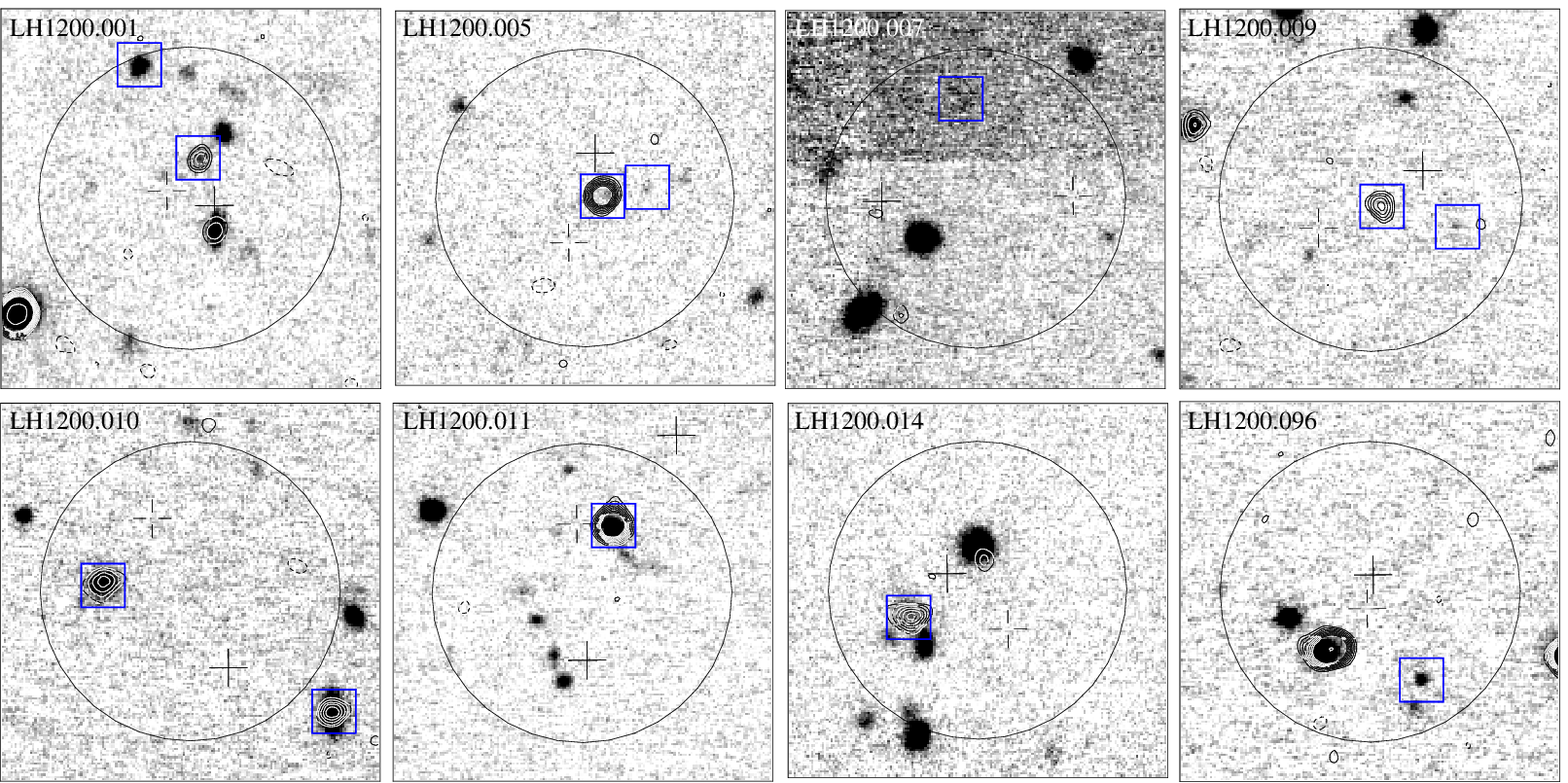,angle=0,width=6.9in}}
\vspace{-0.3cm}
\noindent{\small\addtolength{\baselineskip}{-3pt}}
\caption{Postage stamps (20\,$\times$\,20\,arcsec) of the fields surrounding
LH\,1200.001, .005, .007, .009, .010, .011, .014 and .096. Near-IR ($K$) data
are shown as a grayscale upon which 1.4-GHz contours are plotted at
$-$3,3,4,5,6,7,8,9,10,20,30,40,50 $\times$ $\sigma$.  Open crosses mark SCUBA
galaxies; solid crosses mark MAMBO galaxies; 8-arcsec-radius circles mark the
average positions; squares mark objects discussed in the text. N is up; E is to
the left.}
\label{niri}
\end{figure*}

\section{Optical/infrared observations}

Divining robust counterparts and redshifts for SMGs is a challenging
process, with fewer than ten published redshifts prior to
2002. Ideally, the process follows the pattern outlined below:

\begin{enumerate}
\item definition of submm sample (e.g.\ Scott et al.\ 2002; Greve et al.\
2004);
\item determination of robust counterparts, using sensible priors and
data of sufficient quality in the radio and optical/IR (e.g.\ I02;
Borys et al.\ 2004);
\item spectroscopy, preferably with an instrument sensitive from the blue
atmospheric cut-off ($\sim$310\,nm) to $\sim$1\,$\mu$m.
\end{enumerate}

\noindent
For the particularly retentive, when the spectroscopic identification
is not for a radio-identified galaxy these should be augmented with:

\begin{enumerate}
\setcounter{enumi}{3}
\item confirmation of redshift in IR via H$\alpha$ or another nebular
line (Ivison et al.\ 2000; Simpson et al.\ 2004; Swinbank et al.\
2004);
\item detection of molecular gas at that redshift via CO (e.g.\ Neri
et al.\ 2003; Greve et al.\ 2005).
\end{enumerate}

We have endevoured to follow this process, although only the last two
of the seven Keck runs described by C05 preceded the arrival of the
MAMBO catalogue, so steps (iii) and (iv) have largely preceeded (i)
and (ii). This should not have introduced any strong bias but it has
inevitably reduced the number of robust identifications with
successful redshift determinations.

\subsection{Imaging with Subaru and Gemini}

To illustrate the precise coverage of our spectroscopic observations,
we exploit $R$-band archival imaging taken with the 8-m Subaru
Telescope using SuprimeCam (3$\sigma$ limit $\sim$ 27.4,
1.5-arcsec-radius aperture). Postage stamps of these data are shown in
Fig.~3, with the radio counterparts highlighted --- those with $P<\rm
0.05$ from \S3.2. For the purposes of cross-identification of
radio/optical counterparts --- i.e.\ to match the radio astrometry ---
the optical image required a shift of $\Delta\alpha=-$0.5\,arcsec and
$\Delta\delta=-$0.4\,arcsec. We report the $R$-band photometry of
potential counterparts in Table~3.

New near-infrared ($K$) imaging (Fig.~4) of eight of the SMGs is also
exploited here, obtained in photometric conditions with seeing
$<$0.6\,arcsec at the 8-m Gemini Observatory, Mauna Kea, using NIRI
(GN/01A/11) with a total integration time of 5.4\,ks per source. The
resulting 3$\sigma$ limit in $K$ is 20.9 for a 1.5-arcsec-radius
aperture. As we have only partial coverage of our fields in $K$, we
only include it in our qualitative discussion of the properties of
potential counterparts in these regions in Appendix A.

\subsection{Spectroscopy with Keck}

%
%
\setcounter{figure}{4}
\begin{figure}
\centerline{\psfig{file=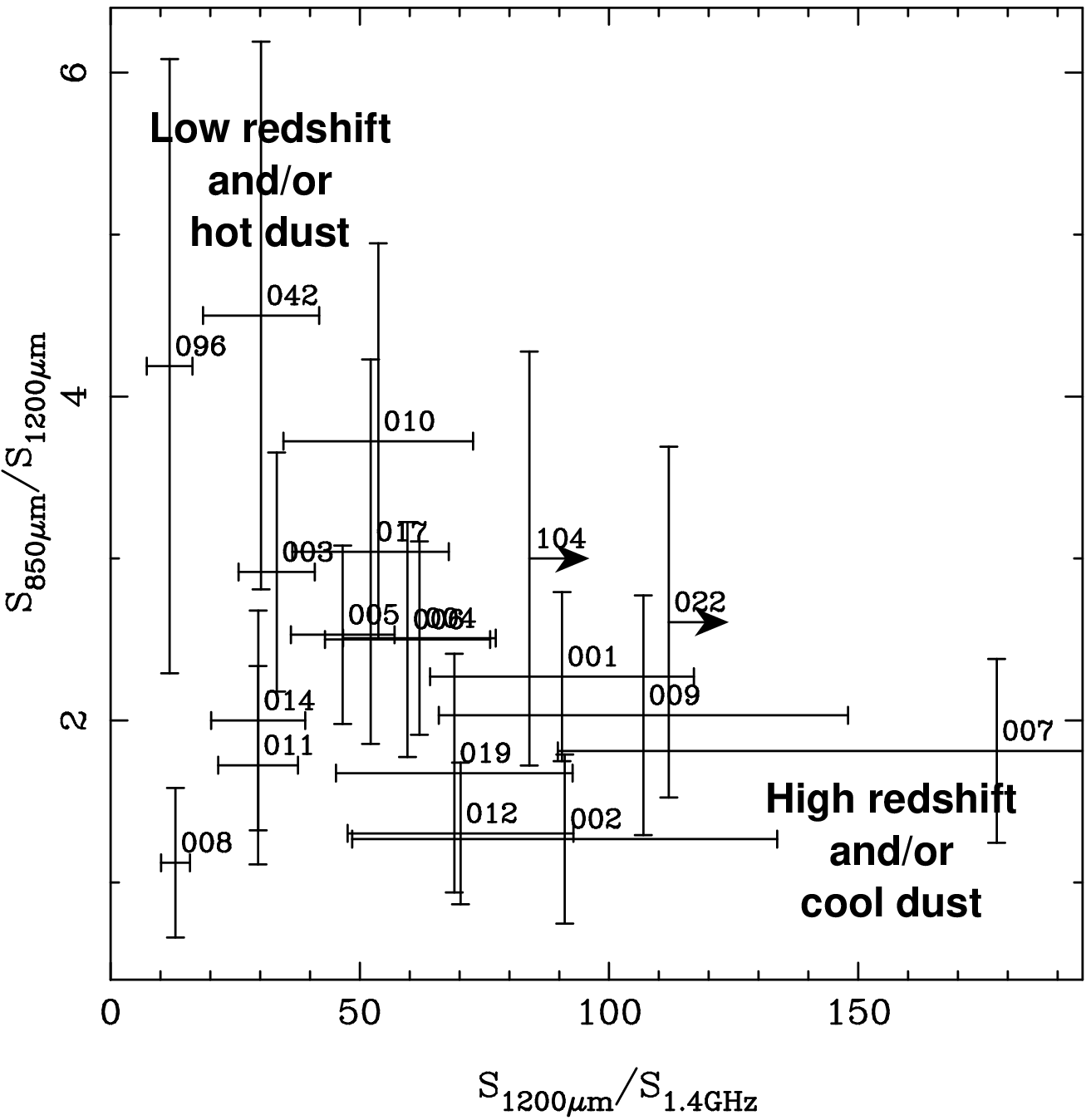,angle=0,width=3.2in}}
\vspace{-0.3cm}
\noindent{\small\addtolength{\baselineskip}{-3pt}}
\caption{0.85-/1.2-mm flux ratio versus the 1.2\,mm/1.4\,GHz
flux ratio, for our robust SMG sample. The plot shows a large
dispersion in both flux ratios, broader than the estimated errors,
indicating a true dispersion in their properties. As we discuss
later, these two flux ratios provide a crude indication of
redshift or the characteristic dust temperature. SMGs in
the upper-left region of the plot are expected to be either at low
redshift or to have a higher characteristic dust temperature;
those SMGs in the lower-right region may be at high redshift or have
cooler dust. The absence of SMGs in the upper-right of the
plot arises due to the lack of very hot, high-redshift sources with
spectra which are consistently steep from 0.85\,mm, through 1.2\,mm,
and out to 1.4\,GHz. To provide the most conservative limits we have
added the radio flux densities where multiple counterparts exist.}
\label{colcol}
\end{figure}

%
%
\setcounter{figure}{5}
\begin{figure}
\centerline{\psfig{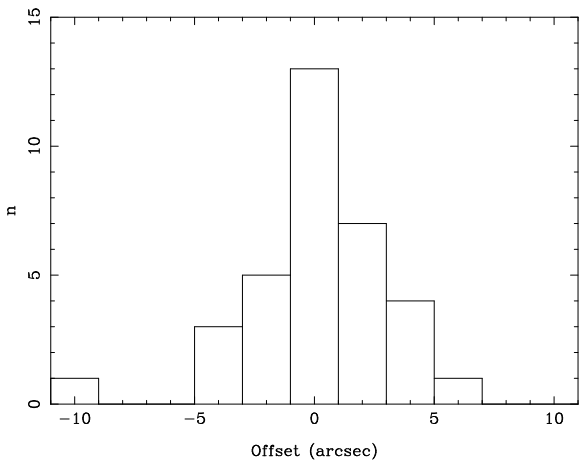}}
\vspace{-0.3cm}
\noindent{\small\addtolength{\baselineskip}{-3pt}}
\caption{Histogram of positional offsets in R.A.\ and Dec.\ between the average
of the 0.85- and 1.2-mm positions and that of the most likely radio
counterpart, negative values corresponding to radio sources SE of a SW--NE line
through the (sub)mm position. A Gaussian fit to the distribution yields a {\sc
fwhm} of 5.2\,$\pm$\,1.2\,arcsec, which translates into an 1-$\sigma$
uncertainty of 2.2\,arcsec for our (sub)mm positions.}
\label{sigma-pos}
\end{figure}

C05 describe spectroscopy of faint radio galaxies and SMGs with the
LRIS spectrograph (Oke et al.\ 1995) on the 10-m Keck-{\sc
i}\footnote{The W.\ M.\ Keck Observatory is operated as a scientific
partnership among the California Institute of Technology, the
University of California and the National Aeronautics and Space
Administration. The Observatory was made possible by the generous
financial support of the W.\ M.\ Keck Foundation.} telescope during
seven runs between 2002 March and 2004 February. Position for the
1.2\,$\times$\,6.5-arcsec slits were chosen without regard for optical
magnitude, guided where possible by radio identifications. In unused
regions of the masks, slits were added for a variety of non-SMG
objects.

For each mask, C05 obtained 3--12 30-min exposures in the red and blue
arms of the spectrograph, with a useful wavelength range of
310-800\,nm. Conditions were usually photometric, with seeing ranging
from 0.7--1.1\,arcsec.

We have searched the catalogue of Lockman Hole slit positions to find
targets from our new SMG catalogue that were observed by C05.  Nine
slits had been placed on the most likely radio counterpart, that with
the lowest $P$ value, identified in \S3: LH\,1200.001, .003, .005,
.006, .008, .010, .011, .042 and .096. For LH\,1200.004, a slit had
been placed on a nearby radio source with $P<\rm 0.05$. In another
five cases, LH\,1200.005, .009, .012, .014 and .042, slits had been
placed on optical/IR galaxies not coincident with radio emission
(additional slits, in the cases of LH\,1200.005 and .042). For
LH\,1200.007, where there is no robust radio counterpart, a slit had
been placed on a nearby X-ray source. For LH\,1200.001, .003, .010 and
.012, additional slits had also been placed on radio sources outside
of our nominal 99.9-per-cent positional confidence regions.

Swinbank et al.\ (2004) presented near-IR spectra for five of our
targets, using NIRSPEC on Keck {\sc ii} during four runs in
2003--04. These spectra utilised a 42\,$\times$\,0.76-arcsec slit
(marked in Fig.\ 3), yielding a resolution of $\sim$1,500 across the
$K$ band. We obtained 2.4-ks $H$ and $K$ spectra of LH\,1200.001
during 2005 March 16 and 17 {\sc ut}, using identical procedures to
those described by Swinbank et al.\ (2004), although in poorer seeing
($\sim$1\,arcsec).

\subsection{Spectroscopy with Gemini}

To augment the spectroscopic data from Keck, we have obtained spectra
from the Gemini\footnote{The Gemini Observatory is operated by the
Association of Universities for Research in Astronomy, Inc., under a
cooperative agreement with the NSF on behalf of the Gemini
partnership: the National Science Foundation (United States), the
Particle Physics and Astronomy Research Council (UK), the National
Research Council (Canada), CONICYT (Chile), the Australian Research
Council (Australia), CNP (Brazil) and CONICET (Argentina).}
Multi-Object Spectrograph (GMOS North -- GN/02B/42) using 1-arcsec
slits, 2 to 5\,arcsec in length.

Data were taken in queue mode, in photometric conditions and
$\ls$0.8-arcsec seeing, using the B600 grating (spectral range:
$\sim$350--650\,nm). The spectral resolution was
$\lambda/\Delta\lambda \sim\rm 1700$ with an output pixel scale of
0.0912\,nm pixel$^{-1}$. Each observation was split into 3-ks
exposures, with a total integration time of 9\,ks.

Five GMOS slits were placed on the most likely radio counterparts for
LH\,1200.005, .009, .010, .014 and .096. For LH\,1200.001, .002, .006,
.009, .010, .012 and .096, slits were placed on optical/IR galaxies
not coincident with radio emission (additional slits, in the cases of
LH\,1200.006,.009, .010 and .096). Slits were placed on radio sources
outside of our nominal 99.9\% positional confidence regions for
LH\,1200.003 and .096.

In Appendix A we discuss our spectroscopic observations of the individual
sources, one by one, referring to the slit positions shown in Fig.~3.

%
%
\setcounter{figure}{6}
\begin{figure}
\centerline{\psfig{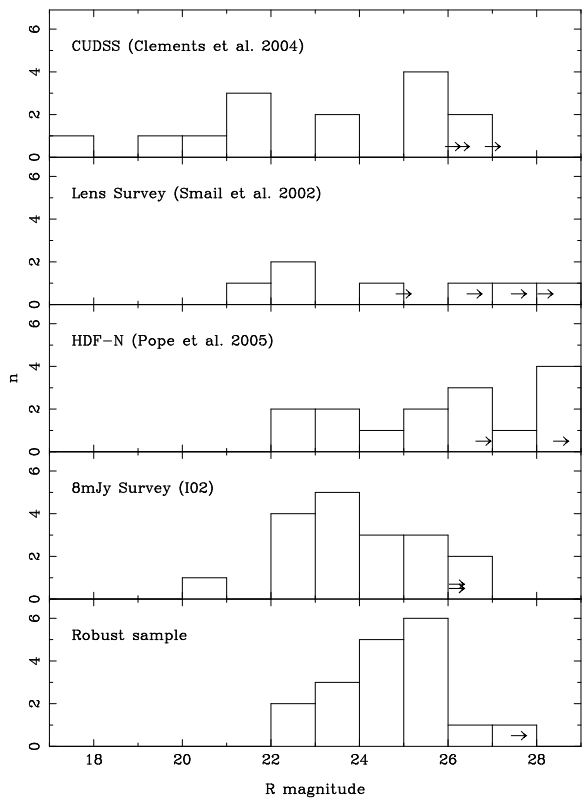}}
\vspace{-0.3cm}
\noindent{\small\addtolength{\baselineskip}{-3pt}}
\caption{Histogram of optical ($R$) magnitudes for radio-identified SMGs in our
sample, compared with those from the surveys by Smail et al.\ (2002), I02,
Clements et al.\ (2004) and Pope et al.\ (2005). Magnitudes have been converted
to a uniform scale (Vega), and the SCUBA Lens Survey data have been corrected
for magnification (Smail et al.\ 2002). We find a broad agreement
between the magnitude distributions with median $R$-band magnitudes of
$R=24.0$--26.2, although our new robust sample shows a smaller
dispersion than many of the other published catalogues.
Where only $I$ or $i_{775}$ magnitudes
are available, the average $(R-I)$ colour (0.75) for 19 radio-identified SMGs
with $R$ and $I$ data has been assumed, somewhat redder than the $R-I=\rm 0.4$
measured for field galaxies (Smail et al.\ 1995). Arrows represent an
SMG for which only a lower limit in $R$ is available.}
\label{histopt}
\end{figure}

%
%
\setcounter{table}{2}
\begin{table*}
\scriptsize
\caption{Redshifts and $R$ magnitudes of MAMBO/SCUBA sources.}
\vspace{0.2cm}
\begin{center}
\begin{tabular}{cclccl}
Name$^a$&$R$ mag$^b$&Magnitude-related notes&$z^c$&Spectrometer&Redshift-related notes\\
&&&&&\\
LH\,1200.001&{\bf 25.67\,$\pm$\,0.08}&NNW radio id                          &{\bf 2.94}&GMOS&Optical galaxy to NW: Ly$\alpha$, N\,{\sc v}, Si\,{\sc ii}\\
            &{\bf 24.34\,$\pm$\,0.05}&SSW radio id                          &3.04?     &LRIS&Ly$\alpha$, N\,{\sc v}?; AGN?\\
LH\,1200.002&     24.83\,$\pm$\,0.04 &Tentative radio id                    &          &GMOS&$z=\rm 0.47$ for nearby optical galaxy\\
LH\,1200.003&{\bf 24.04\,$\pm$\,0.03}&Central radio id                      &{\bf 0.53}&LRIS&\\
            &{\bf 25.40\,$\pm$\,0.12}&S radio id                            &{\bf 1.48}&LRIS&\\
            &                        &                                      &2.43      &GMOS, LRIS&NNW radio id\\
LH\,1200.004&{\bf 26.14\,$\pm$\,0.18}&Central radio id                      &          &LRIS&\\
	    &{\bf 25.22\,$\pm$\,0.07}&S radio id                            &          &LRIS&line at 558nm?\\
LH\,1200.005&{\bf 25.79\,$\pm$\,0.14}&Central radio id; ERO                 &2.15?     &LRIS, NIRSPEC&ERO, 3$''$ to W to radio source, H$\alpha$?\\
LH\,1200.006&{\bf 25.00\,$\pm$\,0.06}&SSE radio id                          &2.14?     &LRIS, NIRSPEC&IS abs; starburst\\
(LH\,1200.007)&3$\sigma>\rm 27.4$      &X-ray source, 21.76\,$\pm$\,0.01      &          &LRIS&$z$ = 0.715 for X-ray source\\
LH\,1200.008&{\bf 22.01\,$\pm$\,0.01}&Central radio id                      &{\bf 1.21}&LRIS&IS abs; starburst\\
LH\,1200.009&{\bf 3\,$\sigma\,>$\,27.4}&Central radio id                    &1.96?     &LRIS&Ly$\alpha$, C\,{\sc iv}; starburst (opt gal 4$''$ SWW of radio)\\
LH\,1200.010&{\bf 22.78\,$\pm$\,0.02}&Confused region                       &{\bf 2.61}&LRIS, GMOS, NIRSPEC&Ly$\alpha$, N\,{\sc v}, Si\,{\sc ii}, H$\alpha$; starburst\\
LH\,1200.011&{\bf 23.68\,$\pm$\,0.03}&NNW radio id                          &{\bf 2.24}&LRIS, NIRSPEC&Ly$\alpha$ abs, C\,{\sc iv} abs, H$\alpha$; starburst\\
LH\,1200.012&{\bf 24.28\,$\pm$\,0.03}&Confused region                       &{\bf 2.66}&LRIS, NIRSPEC&H$\alpha$; see \S5\\
LH\,1200.014&{\bf 23.05\,$\pm$\,0.02}&Confused region                       &{\bf 0.69}&LRIS&[O\,{\sc ii}], Ca HK (2$''$ SSW of brightest radio id)\\
LH\,1200.017&---                     &Diffraction spike                     &          &    &\\
(LH\,1200.019)&3$\sigma>\rm 27.4$      &Brightest object 24.50\,$\pm$\,0.05   &          &    &\\
	    &     23.86\,$\pm$\,0.02 &N radio id                            &          &    &\\
(LH\,1200.022)&3$\sigma>\rm 27.4$      &Brightest object 21.20\,$\pm$\,0.01   &          &    &\\
LH\,1200.042&{\bf 25.17\,$\pm$\,0.07}&WSW radio id                          &{\bf 1.85}&LRIS&IS abs\\
LH\,1200.096&{\bf 24.90\,$\pm$\,0.07}&SE radio id                           &{\bf 1.15}&LRIS&Starburst\\
(LH\,1200.104)&---                     &Diffraction spike                     &          &    &\\
\end{tabular}
\end{center}

\noindent
Notes:
$a)$ Sources in parentheses lack robust radio identifications.\\
$b)$ Values in bold are for sources identified in the radio ($P<\rm 0.05$).\\
$c)$ Values in bold are for redshifts that we consider to be most robust.

\label{speccat}
\end{table*}

\section{Discussion}

\subsection{The distribution of optical magnitudes}

Table~3 lists the {\sc mag\_best} magnitudes of our sample in $R$,
measured using {\sc sextractor}. Fig.~7 shows a histogram of $R$
magnitudes compared with those from the surveys by Smail et al.\
(2002), I02, Clements et al.\ (2004) and Pope et al.\ (2005).  The
various surveys broadly agree, with median $R$-band magnitudes ranging
between $R\sim\rm 24$--26 and a wide range in magnitude within each
sample. The Clements et al.\ (2004) sample has a bright tail which is
less evident in the other surveys. Our sample appears to show the
smallest dispersion, although the presence of upper-limits in all five
samples make this statement hard to quantify.

Nevertheless, from this comparison we can state that the bright,
radio-identified SMG population, spanning barely a magnitude in submm
flux density, covers over 10 orders of magnitude in rest-frame
ultraviolet flux. The optical magnitudes of the radio-identified SMGs
discussed in this paper span around $R=\rm 22$--26, with a faint
tail. The median magnitude is $R=\rm 25.0$, comparable to the sample
analysed by C05, $R=\rm 25.2$, and $\sim$0.6\,mag fainter than the
spectroscopically-identified sample of C05.  As expected, a similar
bias in the spectroscopically identified subset exists within our own
sample, which is $\sim$0.9\,mag brighter than our complete catalogue.

%
%
\setcounter{figure}{7}
\begin{figure}
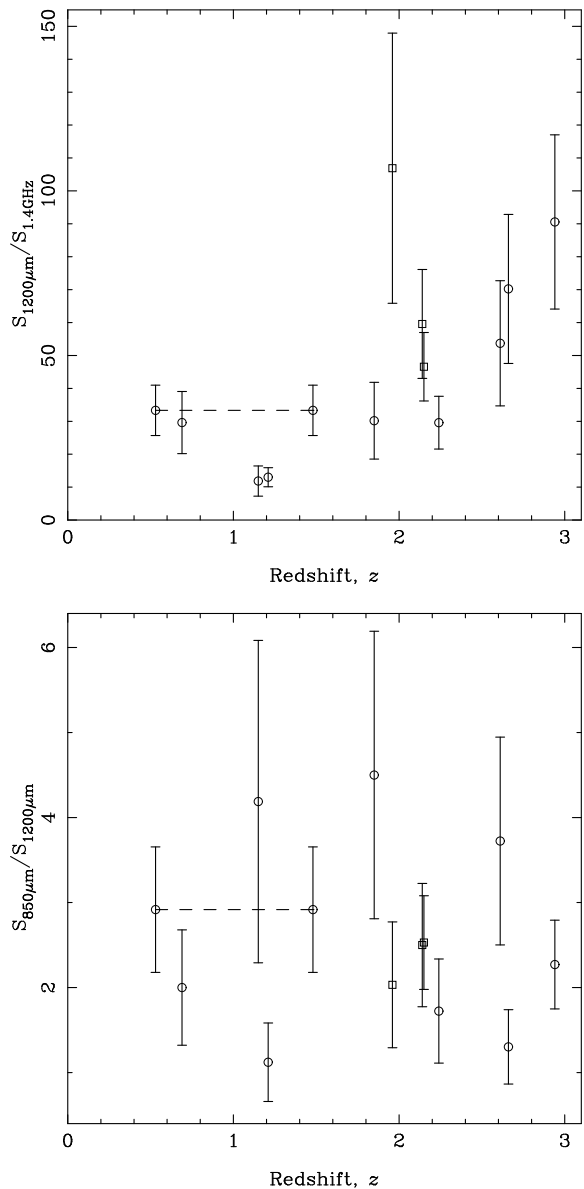

\centerline{\psfig{file=f8a.eps,angle=270,width=3.0in}}
\vspace{0.3cm}
\centerline{\psfig{file=f8b.eps,angle=270,width=3.0in}}
\vspace{-0.3cm}
\noindent{\small\addtolength{\baselineskip}{-3pt}}
\caption{The distribution of flux ratios versus redshift for our
robust (circles) and plausible (squares) redshift identifications.
The upper panel shows the flux ratio between 1.2\,mm and 1.4\,GHz
versus redshift, with a clear trend for higher ratios at higher
redshifts as expected from SED modelling (see Carilli \& Yun 1999;
Greve et al.\ 2004). The lower panel shows the ratio of 0.85-mm
fluxes to those measured at 1.2\,mm, again against redshift. The
scatter in this panel shows that much more precise (sub)mm flux
density measurements will be needed before this ratio can be used to
investigate the redshifts or dust temperatures of SMGs (cf.\ Eales et
al.\ 2003).  Dashed lines link the two redshifts possible for
LH\,1200.003.}
\label{keck1}
\end{figure}

\subsection{The distribution of spectroscopic redshifts}

The redshift distribution determined here inevitably suffers from
spectroscopic incompleteness, but our sample should be otherwise
unbiased. Concentrating on cases where our spectroscopy failed to
secure a robust redshift the failures tend to coincide with the
faintest optical counterparts, $R>\rm 25$, as one might expect.
Looking at the long-wavelength flux ratios (Fig.~5) for those targets
where we failed to obtain identifications or redshifts, e.g.\
LH1200.002 or .004, we see no indication that their (sub)mm/radio
photometric properties differ from those of the sample as a whole.  It
seems unlikely therefore that they lie at substantially higher
redshifts than the subset of SMGs for which we have obtained
redshifts. The exceptions are LH1200.022, and especially LH1200.007,
which have relatively low 0.85-/1.2-mm flux density ratios and fairly
high 1.2-mm/1.4-GHz flux density ratios. This suggests they may lie at
high redshift, or be particularly cold. However, as discussed in
\S3.2, these two sources are not particularly secure (although that
may simply reflect their general faintness in all bands due to their
high redshifts). Reliable identification of these two sources is
therefore only achievable through higher resolution (interferometric)
(sub)mm observations (e.g.\ Lutz et al.\ 2001).

For the six SMGs where we have a single radio identification (and
hence an unambigious counterpart) and for which we have measured
robust redshifts (Table~3), we determine a median of $z=\rm 2.05\pm
0.41$ (where the scatter is estimated from bootstrap re-sampling).
This rises slightly to $z=\rm 2.14\pm 0.27$ if we include the four
sources with robust redshifts for at least one radio counterpart and
the five SMGs where we have less secure redshifts. These figures are
slightly lower than, but entirely consistent with, the spectroscopic
redshift distribution determined by C05 ({\em \={z}} = 2.2). This
suggests that the statistical properties of the C05 sample has not
been strongly biased by the modest significance $>$3$\sigma$) of some of
the submm data used in their analysis.

The median redshift we derive is in reasonable agreement with that
predicted by the {\sc Galform} semi-analytic model (Baugh et al.\
2005), which gives a median redshift of $z=\rm 2.1$ at our flux limit,
with a quartile range of $\pm$0.9 and only 10--20 per cent of the
population at $z\geq\rm 3$. This again suggests that much of the
activity in the bright SMG population is amenable to study via the
precise positions derived from their radio counterparts. This provides
a much-needed route to identify the true far-IR-luminous source within
these frequently morphologically complex and crowded fields.

We illustrate in Fig.~8 the distribution of 1.2-mm/1.4-GHz and
0.85-/1.2-mm flux ratios for our sample versus our spectroscopic
redshifts. The former show a trend to higher flux ratios at higher
redshifts, in line with predictions from SED modelling (e.g.\ Carilli
\& Yun 1999), although with a large scatter; the latter flux ratio,
however, is essentially a scatter plot. We conclude that significantly
more reliable flux measurements (in terms of both absolute calibration
and overall SNR) will be needed to use the 0.85-/1.2-mm flux ratio for
astrophysical analysis.

At face value, both our spectroscopic redshift distribution and that
of C05 are inconsistent with the redshift distribution ({\em \={z}}
$>$ 3) claimed by Eales et al.\ (2003) using the $S_{\rm 850\mu
m}/S_{\rm 1200\mu m}$ flux density ratio (Fig.~8). The Eales et al.\
sample contained 23 sources with $S_{\rm 1200\mu m}$ = 3.1--6.5\,mJy
(cf.\ 19 sources, $S_{\rm 1200\mu m}$ = 1.6--5.7\,mJy here), with a
radio-detected fraction of 73 per cent (cf.\ 79 per cent here). Thus,
the two samples differ only in that the Eales et al.\ photometry
targets were selected above 4\,$\sigma$ at 1200\,$\mu$m rather than
above 5\,$\sigma$ at a combination of 850 and 1200\,$\mu$m. We believe
the disagreement in the mean redshifts derived by these studies most
likely results from the fact that current photometric measurements in
the mm and submm wavebands are insufficiently precise for typical SMGs
to allow them to be used as a reliable redshift indicator,
particularly when some data do not comprise fully sampled images.  The
disagreement may have been compounded by the lower significance of the
Eales et al.\ targets, and the dual-wavelength extraction performed
here could exclude the most distant starbursts (though we note that
the radio-detection of a substantial number of the Eales et al.\
targets is hard to reconcile with the high median redshift claimed for
that sample; indeed, the radio-based estimates given in Eales et al.,
with {\em \={z}} = 2.35, are more consistent with the spectroscopic
results).

\section{Conclusions}

We have developed and applied a dual-survey extraction technique to
SCUBA and MAMBO images of the Lockman Hole, resulting in a robust
sample of 19 SMGs.  Of these, 15 are detected securely by our deep
radio imaging. Those undetected at 1.4\,GHz can be explained by a
combination of contamination by spurious sources (10 per cent) and the
large observed scatter in radio flux densities, which is probably due
to a significant range in dust temperature.

We determine 15 spectroscopic redshifts, of which we consider ten to
be secure.  The resulting redshift distribution ({\em \={z}} = 2.14
for the full spectroscopic sample) is consistent with that determined
for a much larger sample by C05 ({\em \={z}} = 2.2). Our results thus
support the conclusions of C05, who modelled their incompleteness and
estimated only a small shift ($\Delta z = +0.1$) in the median
redshift as a result. Those galaxies for which our spectroscopy failed
to determine redshifts are usually optically faint, $R>\rm 25$, where
the sample ranges from $R=\rm 22$ to $>\rm 26$ with a median of 25.0.

From the radio detections, and the spectroscopy, is seems unlikely
that a significant fraction of bright SMGs lie at very high redshift
($z>\rm 3$) and we conclude that the bright SMG population is readily
amenable to study via radio-selected samples, down to a 850-$\mu$m
flux density limit of $\sim$7\,mJy. "We note, however, that the
dual-wavelength extraction performed here could potentially bias our
sample against very-high-redshift sources which would only be detected
at 1.2mm.

An analysis of separations between SMGs and their radio counterparts
has allowed us to re-calibrate the rule-of-thumb relationship between
positional accuracy, beam size and significance. The most secure SMGs,
at $\sim$10$\sigma$, representative of those expected in upcoming,
confusion-limited, wide-field (tens of square degrees) 0.85-mm surveys
using SCUBA-2 (Audley et al.\ 2004), with lower significance,
simultaneous detections at 0.45\,mm, will be located with a
uncertainty of $\sigma_r \rm \sim 1$\,arcsec, at which level the
precision of the telescope pointing and SCUBA-2 flat-field may become
important contributors to the positional error budget. Assuming these
sources of uncertainty can be minimised, the current requirement for
deep radio coverage to identify counterparts and enable follow-up
spectroscopy may not be as urgent, particularly if
3\,$\times$\,3-arcsec deployable integral-field units are employed for
spectroscopic follow up (e.g.\ KMOS --- Sharples et al.\ 2003, 2004).

\section*{Acknowledgements}

IS acknowledges support from the Royal Society. AWB acknowledges
support from NSF grant AST-0205937, the Research Corporation and the
Alfred P.\ Sloan Foundation.

%
%
\section*{Appendix A}

Notes on our spectroscopic observations of the individual
sources, referring to the slit positions shown in Fig.~3.

%
%
\setcounter{figure}{0}
\begin{figure*}
\centerline{\psfig{file=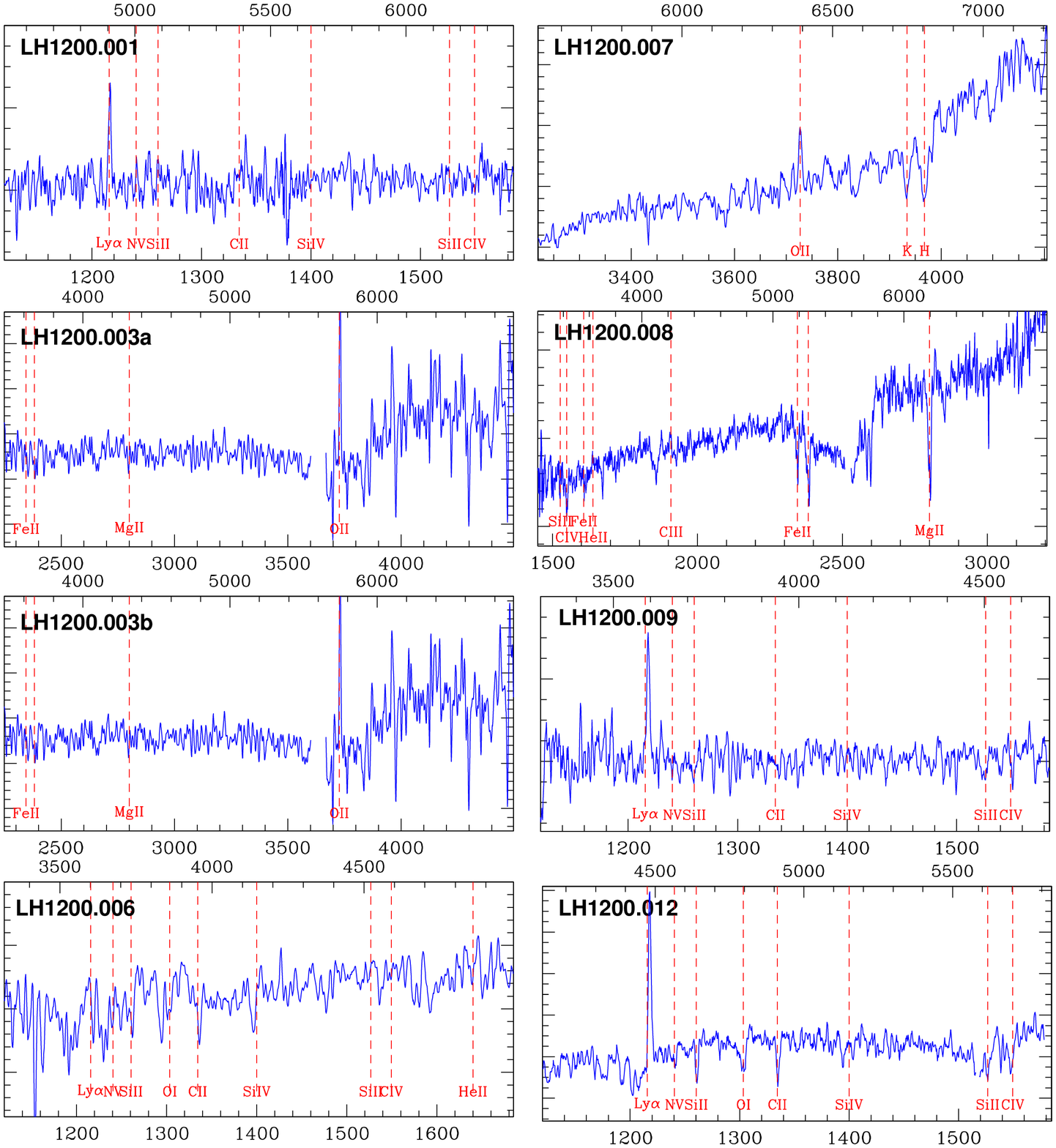,angle=0,width=6.9in}}
\vspace{-0.3cm}
\noindent{\small\addtolength{\baselineskip}{-3pt}}
\caption{Rest-frame UV/optical LRIS spectra of the counterparts discussed in
the text with rest-frame and observed wavelength scales below and above,
respectively, and line identifications shown.}
\label{keck1}
\end{figure*}

%
%
\setcounter{figure}{0}
\begin{figure*}
\centerline{\psfig{file=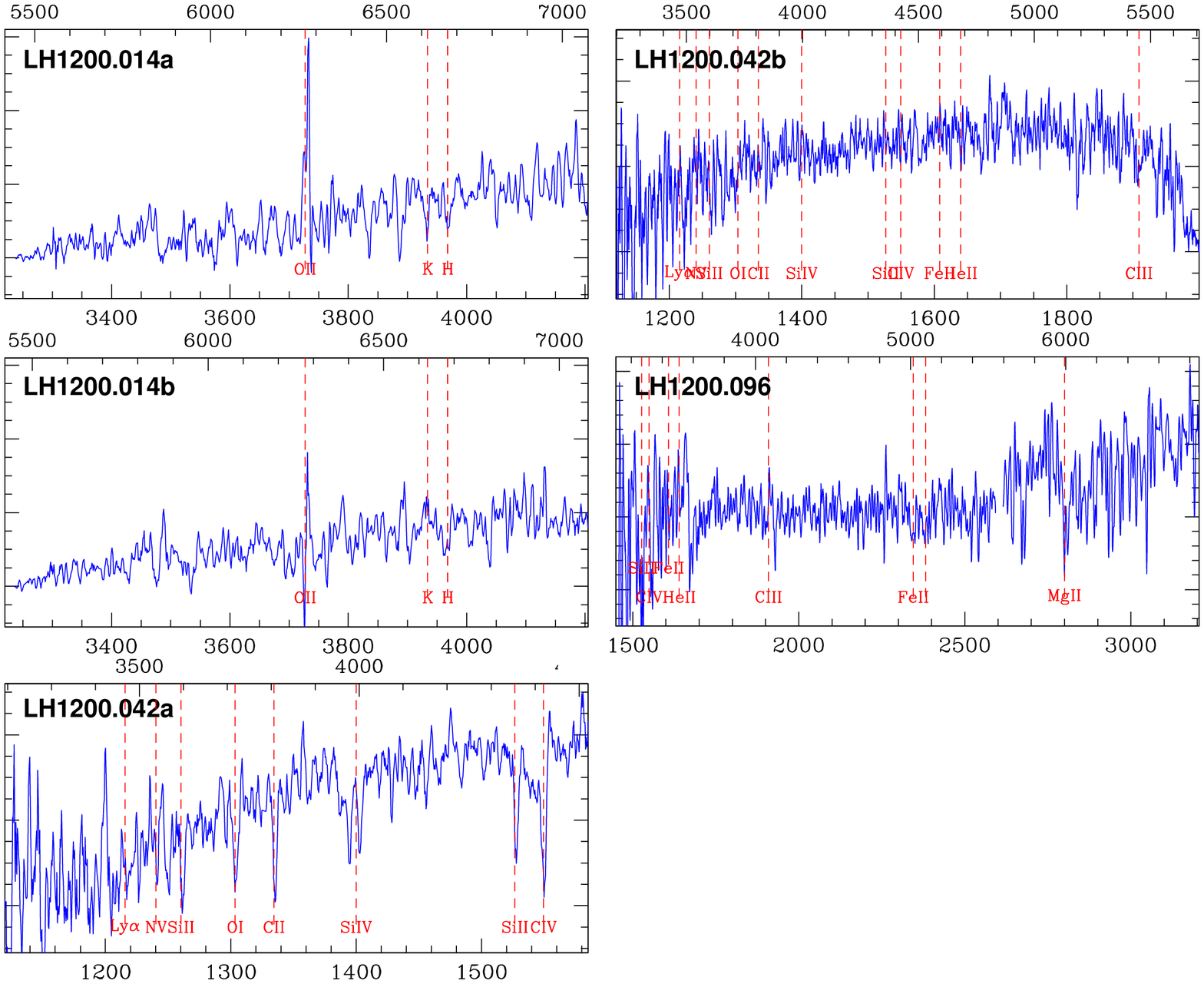,angle=0,width=6.9in}}
\vspace{-0.3cm}
\noindent{\small\addtolength{\baselineskip}{-3pt}}
\caption{continued...}
\label{keck2}
\end{figure*}

%
%
\setcounter{figure}{1}
\begin{figure*}
\centerline{\psfig{file=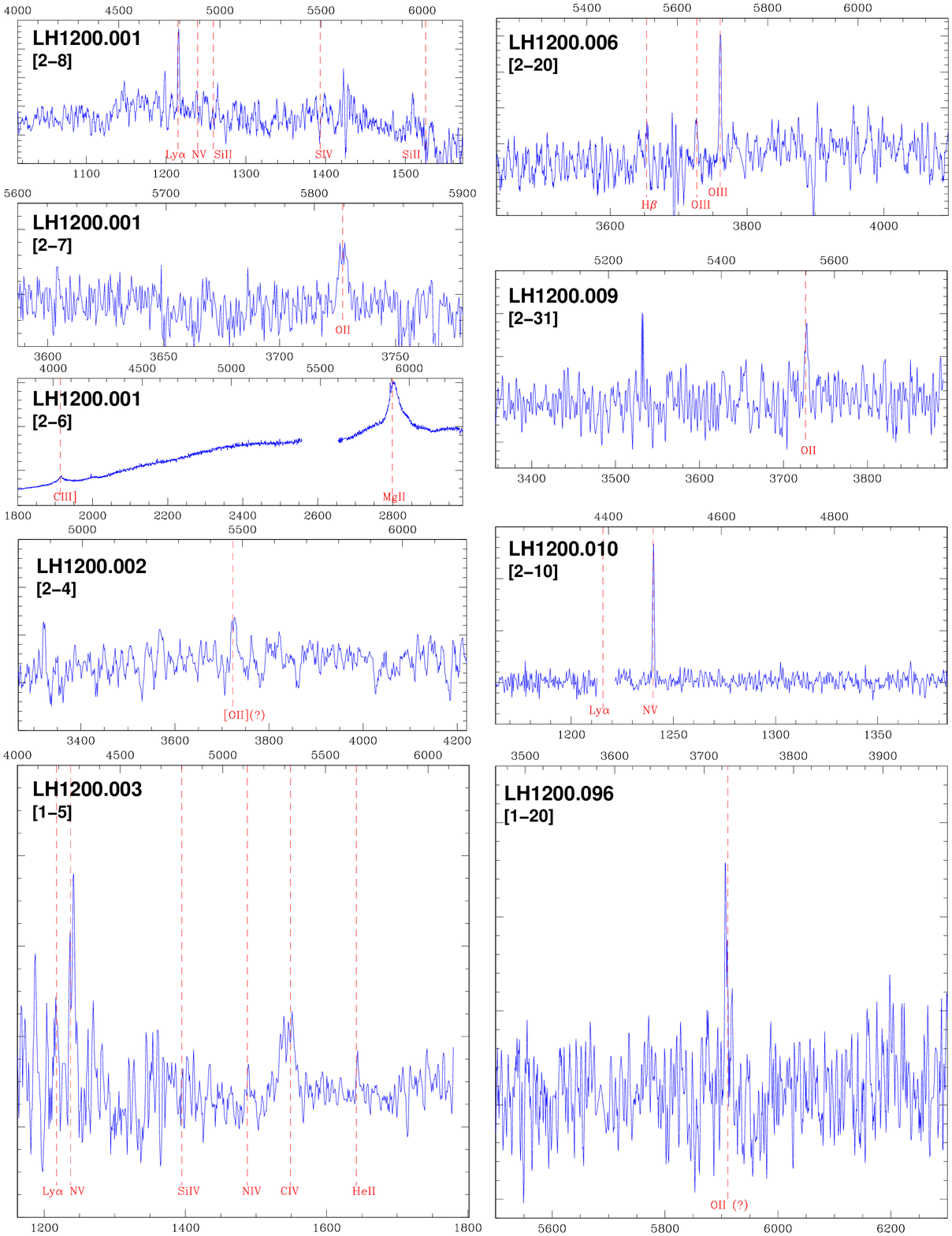,angle=0,width=6.9in}}
\vspace{-0.3cm}
\noindent{\small\addtolength{\baselineskip}{-3pt}}
\caption{Rest-frame UV/optical GMOS spectra of the counterparts discussed in
the text with rest-frame and observed wavelength scales below and above,
respectively. Line, mask and slit identifications are shown.}
\label{gmos}
\end{figure*}

\vspace*{1mm} \noindent {\bf LH\,1200.001:} A complex field. LRIS
slits on several masks had been placed on a faint $R$-band source
coincident with the most probable radio counterpart ($P=\rm
0.024$). Its spectrum, shown in Fig.~A1, is consistent with a redshift
of 3.036, with Ly\,$\alpha$ and hints of weak N\,{\sc v} and several
other lines. We searched for [O\,{\sc ii}] and [O\,{\sc iii}] using
NIRSPEC (\S4.1), yet we were not able to confirm or rule out $z=\rm
3.036$ with confidence. A search for CO(3--2) at IRAM was unsuccessful
(Greve et al.\ 2005), though this could have been due to insufficient
velocity coverage. The second radio counterpart, 4\,arcsec to the
north and also with $P<\rm 0.05$, is associated with an ERO
(Fig.~4). The NIRSPEC longslit was positioned to cover this galaxy; no
lines were evident. Its closest neighbour was observed with GMOS
(Fig.~A2; [2--8]) and found to have emission lines (Ly\,$\alpha$,
possibly N\,{\sc v} and Si\,{\sc ii}) consistent with $z\rm=2.943$
(the apparent continuum blueward of Ly\,$\alpha$ is merely an artifact
of the reduction process). At the edge of the same slit, near the
northern radio position, there is tentative evidence for line emission
from Ly\,$\alpha$, N\,{\sc v} and possibly C\,{\sc iv} at
$z\rm=2.88$. The optically bright radio source 11\,arcsec SE of the
SMG centroid (Fig.~A2; [2--6]) is an AGN, with broad C\,{\sc iii}] and
Mg\,{\sc ii} emission lines, this time at $z=\rm 1.11$. The optical
galaxy 8\,arcsec SSE of the SMG centroid (Fig.~A2; [2--7]) has [O\,{\sc
ii}] at $z=\rm 0.56$. A further ERO, overlooked by I02, and detected
at 3.5$\sigma$ in our smoothed radio image, lies 7.5 arcsec NNE of the
SMG centroid (Fig.~4).  We conclude that the submm emission likely
originates from the two central radio components, which most probably
lie at $z\sim\rm 3$.

\noindent {\bf LH\,1200.002:} Two galaxies in this field were targeted
spectroscopically with GMOS. One target, slit [2--5], lies close to
the faint radio emission near the SMG centroid, but we were unable to
identify its redshift. The other target --- probably unassociated with
the SMG --- has a faint emission line which, if [O\,{\sc ii}],
indicates $z=\rm 0.469$ (Fig.~A2; [2--4]).

\noindent {\bf LH\,1200.003:} Another complex field. LH\,1200.003 may
be a blend of 2--3 SMGs, with MAMBO resolving the source into two, and
the VLA resolving three faint radio sources (one marginally outside
our radio search region). Slits had been placed on all likely
counterparts. The radio source to the SE, targeted by slits on five
LRIS masks at various position angles, is a starburst at $z=\rm
1.482$. The most probable (central) radio identification is associated
with a galaxy at $z=\rm 0.526$ (Fig.~A1). The galaxy to the NNW ---
possibly associated with faint mm emission, LH\,1200.213 in Greve et
al.\ (2004) --- was targeted spectroscopically using LRIS and GMOS and
is an AGN at $z=\rm 2.43$, with broad C\,{\sc iv} emission, as well as
narrower Ly\,$\alpha$, He\,{\sc ii}, N\,{\sc iv} and N\,{\sc v}
(Fig.~A2; [1--5]).

\noindent {\bf LH\,1200.004:} From two possibilities, an LRIS slit had
been placed on the galaxy with the marginally higher $P$ value (0.044
versus 0.033). The galaxy has a weak line in its spectrum at 558\,nm,
possibly Ly\,$\alpha$, but we were unable to determine the redshift
reliably.

\noindent {\bf LH\,1200.005:} LRIS slits were placed on a faint
optical galaxy $\sim$3 arcsec west of the obvious radio
counterpart. The galaxy is estimated to lie at $z=\rm 2.148$ from
absorption lines in its optical spectrum (C05), a value tentatively
confirmed via H$\alpha$ in the $K$ band, although NIRSPEC
slit-rotation problems mean the line cannot be recovered in the final
spectrum or image. Although optically faint (Fig.~3, $R=\rm 25.78\pm
0.14$), the extremely red object (ERO) described by Lutz et al.\
(2001), seen in Fig.~4 at the radio position, was targeted using LRIS
and GMOS. No redshift was forthcoming. The likelihood of finding a
$z\sim\rm 2$ galaxy so close to the SMG centroid is slim, so the two
may well be associated, but we regard the redshift as tentative until
H$\alpha$ is detected unambiguously.

\noindent {\bf LH\,1200.006:} LRIS slits had been placed on by far the
most probable radio counterpart in the region ($P=\rm 0.033$). Its
optical spectrum was identified by C05 as that of a starburst at
$z=\rm 2.142$ (Fig.~A1). This position was also observed by Swinbank et
al.\ (2004) using NIRSPEC, although only [N\,{\sc ii}] would have been
accessible. Since the one-dimensional optical spectrum is not wholly
convincing, this redshift cannot be relied upon absolutely. The
optically bright object to the NW was targeted by GMOS: a foreground
galaxy ($z=\rm 0.14$) with H$\beta$ and [O\,{\sc iii}] emission lines
evident (Fig.~A2; [2--20]).

\noindent {\bf LH\,1200.007:} There is no secure radio counterpart. An
LRIS slit was placed on a nearby {\em XMM-Newton} X-ray source, which
is marginally detected in our smoothed radio image. It lies at $z=\rm
0.715$ (Fig.~A1), and may contribute submm flux to what could well be a
blended submm source. The faint, red galaxy to the north, barely seen
in a noisy part of our $K$ image (Fig.~4), was not targeted
spectroscopically.

\noindent {\bf LH\,1200.008:} An LRIS slit had been placed on by far
the most probable radio counterpart in the region ($P=\rm 0.0004$), an
$R=\rm 22.0$ galaxy amongst a dense ensemble of fainter objects
(Fig.~3), with absorption lines in its spectrum corresponding to
$z=\rm 1.212$ (Fig.~A1), with C\,{\sc iii}] and He\,{\sc ii} weakly in
emission.

\noindent {\bf LH\,1200.009:} GMOS slits were placed on the radio
counterpart and on an optically bright galaxy to the NNW. An
unambiguous redshift could not be determined for either, although the
NNW galaxy may show faint [O\,{\sc ii}] at $z=\rm 0.489$ (Fig.~A2;
[2--31]). LRIS slits had been placed on a faint $R$-band galaxy
$\sim$6\,arcsec SSW of the obvious radio counterpart. This was due to
an positional offset in an earlier version of the Greve et al.\ MAMBO
catalogue. The galaxy 3\,arcsec NW of the slit centre, 4\,arcsec SWW
of the radio source, appears to be a Ly$\alpha$ emitter at $z=\rm
1.956$ (Fig.~A1), though no other lines are seen. As with LH1200.005,
the likelihood of finding a $z\sim\rm 2$ galaxy so close to the submm
centroid is low, so it may be associated with the radio source and the
SMG. The position of the radio source is blank to $K>\rm 22$ in our
deep Gemini/NIRI imaging (Fig.~4). The Ly$\alpha$ emitter is
detected, barely, in $K$. We note that at least one similarly faint
SMG (SMM\,J14009+0252, $K\sim\rm 21$) has been found at $z\sim\rm 2$ (Smith
et al., in preparation).

\noindent {\bf LH\,1200.010:} LRIS slits had been placed on both of
the radio-identified galaxies in this region with one of them (and
several other radio-quiet galaxies) targeted by GMOS. By far the most
robust counterpart ($P=\rm 0.041$), morphologically complex in $R$
(Fig.~3) but relatively uncluttered in $K$ (Fig.~4), has the spectrum
of a starburst at $z=\rm 2.611$ (C05, SMM\,J105230.73+572209.5),
confirmed in H$\alpha$ by Swinbank et al.\ (2005) and in N\,{\sc v} by
GMOS (Ly\,$\alpha$ falling between chips --- Fig.~A2; [2--10]),
although undetected in CO(3--2) at IRAM (Greve et al.\ 2005). The LRIS
spectrum of the radio-bright disk-like galaxy 10\,arcsec to the SW is
also consistent with this redshift. Nothing was seen at the remaining
GMOS slit positions, [2--11] and [2--12].

\noindent {\bf LH\,1200.011:} An LRIS slit had been placed on by far
the most probable radio counterpart in the region ($P=\rm 0.017$), a
compact source with a 5-arcsec-long tail visible in $R$ and $K$ (Figs
3 \& 4). It has the spectrum of a starburst at $z=\rm 2.239$ (C05,
SMM\,J105158.02+571800.2), confirmed convincingly in H$\alpha$ and
[N\,{\sc ii}] by Swinbank et al.\ (2005). The fainter radio source,
14\,arcsec to the NEE, lies at $z=\rm 1.047$. Simpson et al.\ (2004)
identified a $J$-band feature, presumably a noise spike, as [O\,{\sc
ii}] at $z=\rm 2.12$ (rest-frame 359\,nm for $z=\rm 2.34$), as well as
a continuum break at 1.2\,$\mu$m (consistent with the Balmer break at
either redshift).

\noindent {\bf LH\,1200.012:} A complex field, with the robust radio
identification ($P=\rm 0.012$) lying between 3--4 galaxies visible in
$R$, and on top of a galaxy seen in the {\em Spitzer} 3.6-$\mu$m
imaging of Huang et al.\ (2004). A NS-oriented LRIS slit had been
placed on the brightest of the $R$-band galaxies, $\sim$2\,arcsec SSE,
a starburst at $z=\rm 2.686$ (Fig.~A1; C05,
SMM\,J105155.47+572312.7). The NIRSPEC slit, centred on the same
point, yielded continuum but no strong lines. However, 2\,arcsec NW
along the slit --- where fuzz is visible in $R$, 1\,arcsec SW of the
radio centroid --- there is more red continuum emission a line at
2.405\,$\mu$m, corresponding to H$\alpha$ at $z=\rm 2.664$
(Fig.~A3). We view the likelihood of its association with the SMG as
high. Another LRIS slit was placed on the radio source to the NW,
associated with another {\em Spitzer} galaxy, this time at $z=\rm
1.677$ ($z=\rm 1.681$ in H$\alpha$ according to Swinbank et al.\
2005). GMOS slits --- [1--29] through [1--32] --- were placed on all
of the brightest optical knots, but nothing was seen in the spectra.

\noindent {\bf LH\,1200.014:} An interesting case, with strong
similarities to LH\,1200.010 and .012. The brightest of the two radio
counterparts --- associated with faint $K$-band emission (Fig.~4) ---
is the least likely identification, although by a small margin ($P=\rm
0.030$ versus 0.022). LRIS slits were placed on the brightest $R$-band
galaxy (Fig.~3), just to the WSW, which displays [O\,{\sc ii}] in
emission as well as Ca H/K and several Balmer lines in absorption at
$z=\rm 0.689$ (Fig.~A1 --- LH\,1200.014a; C05,
SMM\,J105200.22+572420.2). Its association with the SMG, and with the
radio-identified galaxies, is plausible, as is the possibility that it
acts as a lens (Chapman et al.\ 2002). The slit passed over the radio
centroid which appears to share the same broad spectral features
(Fig.~A1 --- LH\,1200.014b). A GMOS slit, [1--23], was placed directly
on the radio centroid, but the spectral coverage ($<$550\,nm in this
case) meant we were unable to confirm the presence of [O\,{\sc ii}].

%
%
\setcounter{figure}{2}
\begin{figure}
\centerline{\psfig{file=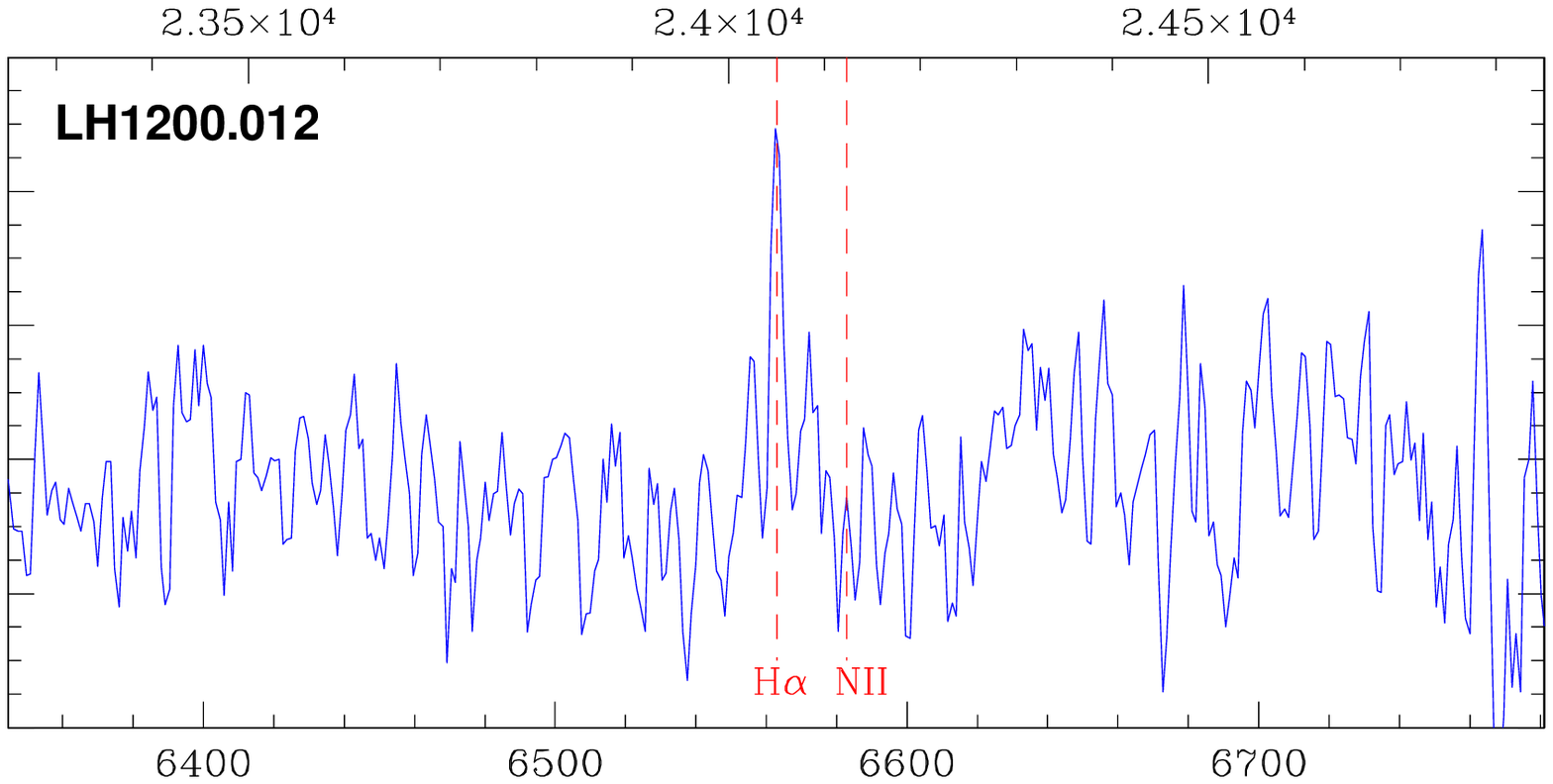,angle=0,width=3.2in}}
\vspace{-0.3cm}
\noindent{\small\addtolength{\baselineskip}{-3pt}}
\caption{NIRSPEC spectrum of LE\,1200.012, extracted from close to the position
of the radio centroid (1\,arcsec SW), with a tentative line detection at
2.405\,$\mu$m assumed here to be H$\alpha$ at $z=\rm 2.664$.}
\label{nirspec}
\end{figure}

\noindent
{\bf LH\,1200.017:} A robust radio identification that was not targeted
spectroscopically.

\noindent
{\bf LH\,1200.019:} There is no obvious identification in this field.

\noindent
{\bf LH\,1200.022:} Again, no obvious identification to target
spectroscopically.
 
\noindent {\bf LH\,1200.042:} LRIS slits had been placed on by far the
most probable radio counterpart in the region ($P=\rm
0.034$). Although optically faint, it has an absorption-line spectrum
consistent with $z=\rm 1.853$ (Fig.~A1 --- LH\,1200.042a). The bright
optical galaxy 3\,arcsec to the SSE, targeted by both GMOS [2--25] and
LRIS, is associated with very faint radio emission and is tentatively
consistent with $z=\rm 1.85$ (Fig.~A1 --- LH\,1200.042b).

\noindent {\bf LH\,1200.096:} Again, LRIS and GMOS slits had been
placed on the most probable radio counterpart ($P=\rm 0.011$), an ERO
(I02) visible out to 24\,$\mu$m (Egami et al.\ 2004), ignoring several
brighter optical galaxies. The LRIS spectrum was classified as a
starburst at $z=\rm 1.147$ by C05 (Fig.~A1 ---
SMM\,J105151.69+572636.0); GMOS saw nothing at this position. This is
the curious SMG discussed by I02, apparently associated with the
steep-spectrum lobe of a radio galaxy, the flat-spectrum core of which
lies to the west, with [O\,{\sc ii}] evident at $z=\rm 0.586$ in its
spectrum (Fig.~A2, [1--20]). Is this system a jet-triggered burst, a
galaxy projected onto an unrelated radio lobe, or a faint, dusty,
ultra-steep-spectrum radio galaxy? Our picture of this system is
muddled and contradictory. To confuse matters further, our NIRI
$K$-band imaging (Fig.~4) reveals another ERO, WSW of the radio
emission, visible out to 8\,$\mu$m in the {\em Spitzer} imaging, and
just missed by GMOS slit [1--19] on a nearby, bluer galaxy.

\noindent {\bf LH\,1200.104:} Despite the lack of radio detections in
the vicinity, a bright {\em Spitzer} counterpart described by Ivison
et al.\ (in preparation) suggests this SMG is not spurious. A
diffraction spike from the nearby star makes identification impossible
in our optical imaging.

\end{document}